\documentclass[twocolumn]{aastex631}

\usepackage{mhchem}

\def\farcs{\hbox{$.\!\!^{\prime\prime}$}} 

\shorttitle{Chemistry in externally FUV irradiated disks in the outskirts of the Orion Nebula}
\shortauthors{Díaz-Berríos et al.}

\graphicspath{{./}{}}

\begin{document}

\title{Chemistry in externally FUV irradiated disks in the outskirts of the Orion Nebula Cluster}

\author[0000-0003-0771-5343]{Javiera K. Díaz-Berríos}
\affiliation{Institute of Astrophysics, Pontificia Universidad Católica de Chile, Vicuña Mackenna 4860, Macul, Región Metropolitana, Chile}
\affiliation{School of Physics and Astronomy, University of Leeds, Leeds, United Kingdom, LS2 9JT}

\author[0000-0003-4784-3040]{Viviana V. Guzmán}
\affiliation{Institute of Astrophysics, Pontificia Universidad Católica de Chile, Vicuña Mackenna 4860, Macul, Región Metropolitana, Chile}
\affiliation{N\'ucleo Milenio de Formaci\'on Planetaria (NPF), Chile}

\author[0000-0001-6078-786X]{Catherine Walsh}
\affiliation{School of Physics and Astronomy, University of Leeds, Leeds, United Kingdom, LS2 9JT}

\author[0000-0001-8798-1347]{Karin I. \"Oberg}
\affiliation{Center for Astrophysics $\mid$ Harvard \& Smithsonian, 60 Garden St., Cambridge, MA 02138, USA}

\author[0000-0003-2076-8001]{L. Ilsedore Cleeves}
\affiliation{Department of Astronomy, University of Virginia, 530 McCormick Rd, Charlottesville, VA 22904}

\author[0000-0002-8546-9531]{Elizabeth Artur de la Villarmois}
\affiliation{Institute of Astrophysics, Pontificia Universidad Católica de Chile, Vicuña Mackenna 4860, Macul, Región Metropolitana, Chile}
\affiliation{N\'ucleo Milenio de Formaci\'on Planetaria (NPF), Chile}
\affiliation{European Southern Observatory, Av. Alonso de Córdova 3107, Vitacura, Santiago, Chile}

\author[0000-0003-2251-0602]{John Carpenter}
\affiliation{Joint ALMA Observatory, Av. Alonso de C\'ordova 3107, Vitacura, Santiago, Chile}

\begin{abstract}
\noindent
Most stars are born in stellar clusters and their protoplanetary disks, which are the birthplaces of planets, can therefore be affected by the radiation of nearby massive stars. 
However, little is known about the chemistry of externally irradiated disks, including whether or not their properties are similar to the so--far better--studied isolated disks. 
Motivated by this question, we present ALMA Band 6 observations of two irradiated Class II protoplanetary disks in the outskirts of the Orion Nebula Cluster (ONC) to explore the chemical composition of disks exposed to (external) FUV radiation fields: the 216--0939 disk and the binary system 253--1536A/B, which are exposed to radiation fields of $10^2-10^3$ times the average interstellar radiation field. We detect lines from CO isotopologues, HCN, H$_2$CO, and C$_2$H toward both protoplanetary disks. Based on the observed disk--integrated line fluxes and flux ratios, we do not find significant differences between isolated and irradiated disks. The observed differences seem to be more closely related to the
different stellar masses than to the external radiation field. This suggests that
these disks are far enough away from the massive Trapezium stars, that their chemistry is no longer affected by external FUV radiation. Additional observations towards lower--mass disks and disks closer to the massive Trapezium stars are required to elucidate the level of external radiation required to make an impact on the chemistry of planet formation in different kinds of disks. 
\end{abstract}


\section{Introduction} \label{sec:intro}
Protoplanetary disks of gas and dust around young stars are the birthplace of planets; hence, understanding the physical and chemical structures of disks is essential to determine the composition of the material that can be incorporated into planets and planetesimals. 

During the last decades, the chemical structure and distribution of numerous molecules have been widely studied in protoplanetary disks \citep[e.g.][]{dutrey_chemistry_2007, pegues_alma_2021, oberg_molecules_2021}. A variety of molecules have been observed and detected using the Atacama Large Millimeter/submillimeter Array (ALMA), the Submillimeter Array (SMA), the Northern Extended Millimeter Array (NOEMA), and the Institute for Radio Astronomy in the Millimeter Range (IRAM) 30m telescope; examples are CO \citep[and isotopologues; e.g.,][]{koerner_rotating_1993,dutrey_chemistry_1997,thi_h2_2001,booth_first_2019}, small organics \citep[C$_2$H, CS, CN, HCN, HNC, H$_2$CO, HCO$^+$, DCO$^+$; e.g.,][]{dutrey_chemistry_1997,van_dishoeck_detection_2003,qi_resolving_2008,oberg_disk_2011,guzman_cyanide_2015,hily-blant_direct_2017,furuya_detection_2022}, and even complex species \citep[CH$_3$OH, CH$_3$CN, HC$_3$N, $c-$C$_3$H$_2$, HCOOH, and CH$_3$OCH$_3$; e.g.,][]{chapillon_chemistry_2012,qi_first_2013,oberg_cometary_2015,walsh_first_2016,favre_first_2018,brunken_major_2022}. These studies have advanced our understanding of the chemistry of planet formation; however, they have focused on disks around isolated stars in low-mass star-formation regions, whereas many stars form in clusters, with low-mass and high-mass stars forming in conjunction \citep{lada_embedded_2003, krumholz_star_2019}. Thus, the chemistry of externally irradiated protoplanetary disks remains poorly constrained observationally.
Following the nomenclature also used in \cite{walsh_molecular_2013}, ``isolated'' refers to a disk irradiated by its central star \textsl{only}, whereas ``irradiated'' refers to a disk illuminated by the central star \textsl{and} the interstellar radiation field (ISRF), which includes at least one nearby massive star.

Most studies of extremely irradiated protoplanetary disks have been focused on the disk mass \citep[e.g.,][]{eisner_protoplanetary_2018, mann_alma_2014, boyden_protoplanetary_2020, boyden_chemical_2023}, radius \citep[e.g.,][]{vicente_size_2005, clarke_photoevaporation_2007, mann_alma_2014, boyden_protoplanetary_2020, boyden_chemical_2023}, evolution \citep[e.g.,][]{champion_herschel_2017, haworth_first_2017}, and lifetime \citep[e.g.,][]{adams_photoevaporation_2004, haworth_proplyds_2021, winter_solution_2019}. 
Theoretical models and observational studies have predicted and discovered, respectively, that disks exposed to external radiation decrease in their mass and size when they are close to the massive stars, due to photoevaporation \citep[e.g.,][]{johnstone_photoevaporation_1998, storzer_photodissociation_1999, matsuyana_viscous_2003, adams_photoevaporation_2004, mann_alma_2014, boyden_protoplanetary_2020, concha-ramirez_evolution_2023, boyden_chemical_2023}. 
This is expected to affect the ability of these disks to form planets. However, photoevaporation is expected to mainly remove gas from the outer disk. This is because dust can grow in size and the subsequent settling/drift lead to a depletion of dust in the surface and outer regions of the disk where external photoevaporation is mainly operating \citep{Facchini2016}. Therefore, rocky planets may still be able to form in the inner disk midplane \citep{adams_photoevaporation_2004, concha-ramirez_evolution_2023}. 

We currently do not know the impact of an external radiation field on the chemistry of a disk. This is important to investigate because isolated disks are not the norm in the Galaxy. Instead, the vast majority of stars form within rich stellar clusters where disks are constantly affected by intense radiation fields from nearby massive stars, and exposed to extreme-ultraviolet (EUV) radiation and far-ultraviolet (FUV) radiation \citep{adams_photoevaporation_2004}, that affects their evolution through time. Our Sun is thought to have been born in such a cluster, so the protosolar nebula would have been irradiated by its massive neighbors \citep{lada_embedded_2003, adams_birth_2010, parker_birth_2020}. Hence, understanding the initial chemical conditions of the protosolar nebula requires an understanding of the chemistry in disks located in massive star-forming regions.

Observing molecular lines in disks that are close to massive stars is challenging, because of their physical size \citep[$R_\mathrm{disk}\sim 10^2-10^3~\mathrm{au}$;][]{mann_alma_2014, boyden_chemical_2023} and the large distances to massive star-formation regions, meaning that they typically span only $\sim0\farcs2-1\farcs2$ on the sky (at the distances of the nearest massive star forming regions). Additionally, many disks in clustered regions are still embedded in the parent molecular cloud, and it can be challenging to disentangle the disk molecular line emission from the cloud molecular line emission. 

A few astrochemical disk models have investigated the effect of strong external FUV fields, characteristic of the environments close to O/B stars, on disk chemistry. These models predict that external FUV sources will significantly impact the thermal and chemical structure of the disk \citep{nguyen_molecular_2002, walsh_molecular_2013, walsh_complex_2014}. Therefore, the conclusions reached for isolated disks may not be directly transferred to disks born near massive stars.
For example, the gas temperature is expected to be significantly higher in the outer disk midplane in externally irradiated disks, which will result in the release of molecules that would usually be frozen out onto dust grains \citep{walsh_molecular_2013}. 
However, these models remain speculative and their predictions are yet to be tested through comparisons with observations.

In this work, we present observations of two protoplanetary disks around pre-main sequence stars located in the outskirts of the Orion Nebula Cluster (ONC), which is the closest ($\sim414$ pc) massive star-forming region \citep{menten_distance_2007, rzaev_binary_2021}. Our aim is to investigate the possible differences between the chemistry of isolated disks and of externally irradiated disks. By studying disks with similar initial conditions to the protosolar nebula, such as externally irradiated disks, we can learn about the formation conditions of planets in the Solar System. The ONC is an excellent laboratory to study these kinds of sources. 
This rich cluster has thousands of stars, but the radiation field is dominated by a single star: $\theta^1$ Ori C \citep{odell_postrefurbishment_1994, smith_new_2005, ricci_hubble_2008}, a young \citep[$\sim 1$ Myr;][]{hillenbrand_stellar_1997}, massive \citep[$45~\mathrm{M}_\odot$;][]{kraus_tracing_2009, rzaev_binary_2021}, O6-type star \citep{odell_which_2017}.
Hubble Space Telescope (HST) observations revealed hundreds of disks in Orion that are externally irradiated, some of them surrounded by a cometary ionization front -- the so-called proplyds \citep{odell_discovery_1993, odell_postrefurbishment_1994, ricci_hubble_2008, eisner_protoplanetary_2018}. More recent ALMA surveys have shown that disks close to the stellar cluster ($\lesssim 0.5$ pc) can have a significant amount of their disk mass removed due to external photoevaporation, as was predicted by \citet{johnstone_photoevaporation_1998} and \citet{storzer_photodissociation_1999}, while disks that are farther away have disk masses that are relatively intact \citep{mann_alma_2014, eisner_protoplanetary_2018,vanTerwisga2019}. However, the chemistry of disks in these less exposed regions could be affected because the FUV radiation, even if it is not as extreme as in the EUV regime, is still higher than the FUV field to which the isolated disks are exposed. 

Section \ref{sec:obs} describes the ALMA observations, and provides information on the sources, including the disk and stellar properties, and the spectroscopic parameters of the molecular lines. Section \ref{sec:results} presents the results of the continuum emission and molecular lines. In Section \ref{sec:discussion} we compare the results of the externally irradiated disks with those for isolated disks. Finally, Section \ref{sec:conclusion} presents our conclusions.

\begin{deluxetable}{@{\extracolsep{0pt}}lcc}
\tablecaption{Stellar and disk properties.}
\tablehead{
    & \colhead{216--0939} &  \colhead{253--1536A/B} 
    }
    \startdata
    \hline  
     RA & $5^{h}35^{m}21.57^{s}$ & $5^{h}35^{m}25.30^{s}$/$5^{h}35^{m}25.23^{s}$ \\
     DEC & $-5^{\circ}9^{\prime}38.9^{\prime\prime}$ & $-5^{\circ}15^{\prime}35.4^{\prime\prime}$/$-5^{\circ}15^{\prime}35.69^{\prime\prime}$\\ 
     $\mathrm{M_\star}$ & 2.17~$\mathrm{M_\odot}$\tablenotemark{a} & 3.5~$\mathrm{M_\odot}$/$>0.2~\mathrm{M}_\odot/\sin^2 i_\mathrm{B}$\\
     Spectral type & K5 & F/M2 \\
     Distance\tablenotemark{b} & 1.59~pc & 0.92~pc \\ 
     Inclination & $32^{\circ}$ & $65^{\circ}$/- \\
     PA & $173^{\circ}$ & $69.7\pm 1.4^{\circ}$/ $136\pm 15^{\circ}$\\
     Disk mass & 46 M$_\mathrm{Jup}$ & 79/30 M$_\mathrm{Jup}$ \\   
     $v_\mathrm{LSR}$ & 10.75~km/s & 10.55/10.85~km/s \\
     FUV field\tablenotemark{c} & $G_0\sim180$ & $G_0\sim500$ \\
    \hline
    \enddata
\label{table:disk-properties}
\tablenotemark{a} 216--0939 could also be a tight equal-mass binary of two $\sim1~\mathrm{M_\star}$ stars (see Section~2).
\tablenotemark{b} Projected distance to $\theta^1$~Ori.
\tablenotemark{c} Based on \emph{Herschel} FIR observations \citep{Pabst2021}.
\tablecomments{The stellar and disk properties were taken from \cite{factor_alma_2017} and \cite{williams_alma_2014} for the 216--0939 and 253--1536A/B system, respectively. }
\end{deluxetable}

\section{Observations} \label{sec:obs}

We have targeted two protoplanetary disk systems located in the Orion Nebula Cluster at a distance of $414$~pc \citep[e.g.,][]{rzaev_binary_2021}. The two systems, 216--0939 and 253--1536A/B, are located in the outskirts of the ONC (projected distance of $> 0.9 $~pc from $\theta^1$ Ori C), where the radiation field is estimated to be $<10^3~\mathrm{G}_0$ \citep[][]{mookerjea_mapping_2003}, where $\mathrm{G}_0$ is the average strength of the integrated ISRF; $\mathrm{G}_0 \approx 1.6\times 10^{-3}$~erg~cm$^{-2}$~s$^{-1}$ \citep{draine_gas_heating_1978}.
The sources have been previously identified and classified with HST \citep[e.g.,][]{smith_new_2005, ricci_hubble_2008, ricci_sub-millimeter_2011}, and have been observed with ALMA as part of a survey that observed the continuum emission in 22 disks near the ONC \citep[e.g.,][]{mann_alma_2014}. 
The three disks have estimated gas masses of the same order of magnitude, ranging between $\sim 30-80 ~\mathrm{M}_\mathrm{Jup}$. The two systems are exposed to different external radiation fields (by almost an order of magnitude), which allows us to investigate for the first time the effect of different external radiation fields on the chemistry. The stellar and disk propeties are summarized in Table~\ref{table:disk-properties}. This section presents the sources, observational details, and data reduction process.

{\bf{Source 216--0939}} is a K5 star, determined spectroscopically by \citet{hillenbrand_stellar_1997}. However, the dynamical mass of $2.17\pm0.07$~M$_\odot$ estimated by \cite{factor_alma_2017} is inconsistent with this spectral type. They propose that the star could instead be a tight equal-mass binary of two 1.1~M$_\odot$ stars. For the purposes of this study we consider both cases. The disk is located at a projected distance of $1.59$ pc from $\theta^1$ Ori C \citep[O6 star;][]{hillenbrand_stellar_1997, factor_alma_2017}, and at $0.8$ pc from $\nu$ Ori \citep[B3V star;][]{terada_discovery_2012}. The disk around 216--0939 (J2000 R.A. $\mathrm{= 05h35m21.57s}$; J2000 DEC. $\mathrm{= -05d~09m~38.9s}$) has an inclination of $32^\circ$, a position angle ($\mathrm{PA}$) of $173^\circ$, a systemic velocity ($v_\mathrm{LSR}$) of $10.75$ km/s \citep{factor_alma_2017}, and a dust disk size of $\sim290~\mathrm{au}$ \citep{mann_massive_2009}.  
This source is one of the most massive protoplanetary disks in the ONC. The gas mass of the disk is estimated to be $\sim 46~ \mathrm{M}_\mathrm{Jup}$ \citep[$\sim 0.04~\mathrm{M}_\odot$;][]{ricci_mm-colors_2011, mann_alma_2014, mann_massive_2009, factor_alma_2017}, computed from the 856$\mu$m dust continuum emission. Additionally, because of its distance to the massive stars, the external FUV radiation field is estimated to be $\mathrm{G}_0\approx 180$, based on Herschel FIR observations \citep{Pabst2021}.
Previous studies by \cite{mann_massive_2009, mann_submillimeter_2010} and \cite{factor_alma_2017} reported the detection of HCO$^+$~($4-3$), CO~($3-2$), and HCN~($4-3$) lines and a weak detection of CS~($7-6$) with an rms noise of $0.41~\mathrm{mJy~beam^{-1}}$ and a synthesized beam of $0\farcs57\times 0\farcs52$, and provided the first view of the molecular content and structure of the disk.

Additionally, HST observations of the 216--0939 star revealed the presence of water ice absorption, concluding that it most likely originated from the surrounding disk \citep{terada_discovery_2012}. The water ice detection is associated with a large silhouette disk, about $\sim 1000~\mathrm{au}$ in diameter and showed that there are regions of the disk that are sufficiently cold to host  substantial ice reservoir \citep{terada_discovery_2012, terada_adaptive_2012}. 

{\bf{Source 253--1536A/B}} is a wide binary system with dynamically estimated stellar masses of $\sim3.5~\mathrm{M}_\odot$ and $\gtrsim0.2~\mathrm{M}_\odot/\sin^2 i$, for the A and B members, respectively \citep{williams_alma_2014}. The lower-mass star, 253--1536B, has a spectral type M2, while a spectral type of F/G has been reported for 253--1536A by \citet{ricci_mm-colors_2011}. The disk of 253--1536A has an inclination of $65^\circ$, and a  position angle ($\mathrm{PA}$) of $69.7^\circ$ \citep{williams_alma_2014}. The inclination is unknown for 253--1536B, but the position angle has been estimated to be $136^\circ$ \citep{williams_alma_2014}. The systemic velocity of the system ($v_\mathrm{LSR}$) is $10.55$ and 10.85~km/s for the A/B members \citep{williams_alma_2014}. 
This system, an ONC proplyd \citep{smith_new_2005}, is located inside the M43 HII region and is associated with a bright proplyd ionization front \citep{williams_alma_2014} and a bipolar jet \citep{smith_new_2005}. 
This binary system is at a projected distance of $0.92$ pc from $\theta^1$ Ori (J2000 R.A. $\mathrm{= 05h35m25.30s/05h35m25.23s}$; J2000 DEC. $\mathrm{= -05d15m35.40s/-05d15m35.69s}$), and its external FUV radiation field is estimated to be $\mathrm{G}_0\sim 500$, based on Herschel FIR observations \citep{Pabst2021}.
The gas masses of the disks are estimated to be $\sim 79$ and $30~\mathrm{M}_\mathrm{Jup}$ for A and B ($\sim 0.08$ and $0.03~\mathrm{M}_\odot$), respectively, computed from the 856$\mu$m dust continuum emission \citep{williams_alma_2014}.
253--1536A has an estimated disk radius of 0\farcs75 equivalent to $\sim300$~au, and is 1\farcs1 (440~au) away from 253--1536B \citep{williams_alma_2014}. No evidence of a larger circumbinary disk around the system has been found \citep{smith_new_2005, mann_massive_2009}. 

The three protoplanetary disks of this study are massive enough ($\ge 30$ M$_\mathrm{Jup}$) to potentially form planets \citep{mann_massive_2009}, providing test cases for studying planet formation under extreme irradiation.

\begin{deluxetable*}{@{\extracolsep{3pt}}cccccc}[t!]
\tablecaption{Spectroscopic parameters of targeted molecular lines.}
\tablewidth{0pt}
\tablehead{
\colhead{Molecule}  & \colhead{Line}    & \colhead{Rest. Freq. (GHz)}  & \colhead{$\log(A_{ij} (\mathrm{s}^{-1}) )$} & \colhead{$E_u$ (K)}   & \colhead{$g_u$}
}
\startdata
    \hline
	\hline
    DCN             &   J~=~$3-2$               &	$217.2385$ &   $-3.3396$   &   $20.85$ &   $21$  \\	
    $c-$C$_3$H$_2$  &   (J$_\mathrm{K_a, K_c}$)~=~$6_{06}-5_{15}$	   &	$217.8221$ &  $-3.2679$    &  $38.61$ &  $13$ \\
    H$_2$CO         &   (J$_\mathrm{K_a, K_c}$)~=~$3_{03}-2_{02}$     &	$218.2222$  &	$-3.5504$	&	$20.96$	&	$7$ \\
    H$_2$CO         &   (J$_\mathrm{K_a, K_c}$)~=~$3_{22}-2_{21}$     &	$218.4756$ &	$-3.8037$	& $68.09$	& $7$\\
    H$_2$CO         &   (J$_\mathrm{K_a, K_c}$)~=~$3_{21}-2_{20}$     &	$218.7601$ &	$-3.8024$	& $68.11$	& $7$\\
    C$^{18}$O	    &   J~=~$2-1$	           &	$219.5604$ &	$-6.2211$	&	$15.81$	&	$5$ \\  
    $^{13}$CO       &	J~=~$2-1$	           &	$220.3987$ &	$-6.2191$	&	$15.87$	&	$5$ \\
	$^{12}$CO	    &	J~=~$2-1$	           &	$230.5380$ &	$-6.1605$	&	$16.60$	&	$5$ \\
    N$_2$D$^+$      &   J~=~$3-2$	           &	$231.3218$ &  $-3.1465$    & $22.20$  & $63$  \\
    $c-$C$_3$H$_2$  &   (J$_\mathrm{K_a, K_c}$)~=~$7_{07}-6_{16}$	   &	$251.3144$ & $-3.0704$  &  $50.67$ & $45$ \\
    $c-$C$_3$H$_2$  &   (J$_\mathrm{K_a, K_c}$)~=~$6_{25}-5_{14}$	   &	$251.5273$ &  $-3.1706$   &   $47.49$ &   $39$\\
	C$_2$H          &   N~=~$3-2$, J~=~$\frac{5}{2}-\frac{3}{2}$, F~=~$3-2$	&	$262.0650$ &	$-4.1521$	&   $25.16$ &   $7$\\
	HCN		        &   J~=~$3-2$               &	$265.8864$ &	 $-3.0766$           &	 $25.52$   	&	$21$    \\
                    &   J~=~$4-3$               &	$354.5055$ &	  $-3.1614$          &	  $42.53$  	&	$27$    \\
	\hline
\enddata
\tablecomments{Molecular data extracted from the CDMS \citep{muller_cologne_2001, muller_cologne_2005, endres_cologne_2016}, JPL \citep{pickett_submillimeter_1998}, and LAMDA molecular \citep{schoier_atomic_2005} catalogues, obtained through \href{http://www.cv.nrao.edu/php/splat/}{www.splatalogue.net} \citep{remijan_splatalogue_2007}.}
\label{table:splatalogue}
\end{deluxetable*}

\subsection{Observations details}
The observations of the 216--0939 and 253--1536A/B protoplanetary disks were obtained with ALMA as part of the Cycle 5 project \#2018.1.01190.S (PI: V. Guzmán). The ALMA Band 6 observations included two spectral settings, at 1.2 and 1.3 mm. The correlator setup was configured with narrow spectral windows targeting different molecular lines. The main targets of the observations were lines from species commonly observed in isolated disks, such as the CO isotopologues, HCN, small carbon chains, H$_2$CO, and deuterated species. Table \ref{table:splatalogue} summarizes the molecular line targets and their spectral properties. 

The Band 6 observations were carried out in August 2018 with baseline lengths spanning between 41 and 3640 m approximately. 
The total on-source time was 82 and 84 minutes per spectral setting, for the 216--0939 and 253--1536 disks, respectively. 
The quasar \mbox{J0423--0120} was observed to calibrate the frequency bandpass and amplitude, and the quasars \mbox{J0529--051} and \mbox{J0607--0834} were observed to calibrate phase temporal variations (see Table \ref{tab:spectral-setting-app}).

Additionally, we used archival ALMA Band 7 observations of the two protoplanetary disks obtained as part of the Cycle 0 project \#2011.0.00028.S (PI: R. Mann), which include the HCO$^+$ ($4-3$) and HCN~($4-3$) lines. More details about these observations can be found in \cite{mann_alma_2014}.

\begin{deluxetable*}{@{\extracolsep{10pt}}cccrccrc}
\tablecaption{Molecular line and continuum image parameters/properties.}
\tablewidth{0pt}
\tablehead{
\colhead{Species} & \multicolumn{1}{c}{Transition}  & \multicolumn{2}{c}{{Beam size}} & \colhead{RMS\tablenotemark{a}} & \multicolumn{2}{c}{{Beam size}} & \colhead{RMS\tablenotemark{a}} \\
      & & \colhead{(arcsec)} & \colhead{(deg)} & \colhead{(mJy/beam)}  & \colhead{(arcsec)} & \colhead{(deg)} & \colhead{(mJy/beam)}
}
\startdata
\hline
& & \multicolumn{3}{c}{\bf{216--0939}} & \multicolumn{3}{c}{\bf{253--1536A/B}}\\
\hline
\hline
Continuum (high-res.) &  & $0.12 \times 0.13$ & $-57.18$ & $0.04$ & $0.09 \times 0.12$ & $82.47$ & $0.08$ \\
Continuum (smooth) & & $0.48 \times 0.51$ & $62.96$ & $0.25$ & $0.47 \times 0.50$ & $64.07$ & $0.78$ \\
\hline
DCN & $3-2$ & $0.48 \times 0.52$ & $67.34$ & $2.37$ & $0.47 \times 0.50$ & $71.95$ & $3.04$ \\
$c-$C$_3$H$_2$ & $6_{06}-5_{15}$ & $0.48 \times 0.52$ & $67.59$ & $2.24$ & $0.47 \times 0.50$ & $70.31$ & $2.88$ \\
H$_2$CO & $3_{22}-2_{21}$ 	&   $0.48 \times 0.52$ & $67.63$  & $2.06$ &   $0.46 \times 0.50$ & $69.54$  & $2.66$  \\
C$^{18}$O & $2-1$	            &   $0.48 \times 0.52$ & $65.97$  & $2.18$ &   $0.47 \times 0.50$ & $66.92$  & $2.83$  \\
$^{13}$CO  & $2-1$	            &   $0.48 \times 0.51$ & $63.96$  & $3.18$ &   $0.47 \times 0.50$ & $64.07$  & $4.14$  \\
$^{12}$CO & $2-1$ 		        &   $0.48 \times 0.51$ & $62.68$  & $2.89$ &   $0.46 \times 0.49$ & $65.95$  & $3.76$  \\
N$_2$D$^+$ & $3-2$    	        &   $0.47 \times 0.51$ & $65.55$  & $2.91$  &   $0.46 \times 0.49$ & $68.78$  & $3.74$    \\
C$_2$H & $3-2$    	            &   $0.45 \times 0.49$ & $-69.88$ & $3.09$ &   $0.45 \times 0.49$ & $-70.97$ & $3.61$  \\
$c-$C$_3$H$_2$ & $7_{07}-6_{16}$ & $0.46 \times 0.50$ & $-72.74$ & $3.44$ & $0.45 \times 0.49$ & $-72.94$ & $3.94$ \\
$c-$C$_3$H$_2$ & $6_{25}-5_{14}$ & $0.46 \times 0.50$ & $-71.89$ & $3.56$ & $0.45 \times 0.49$ & $-73.99$ & $4.11$ \\
HCN & $3-2$		                &   $0.45 \times 0.49$ & $-69.94$ & $3.39$ &   $0.45 \times 0.49$ & $-68.58$ & $4.03$  \\
\hline
\enddata
\tablecomments{Lines were imaged with a spectral resolution of $0.4~\mathrm{km/s}$. 
} 
\tablenotetext{a}{Average rms estimated from the line free channels which correspond to the first $5$ and last $5$ channels in the image cube.}
\label{table:spectroscopic-beam}
\end{deluxetable*}

\subsection{Data reduction}

The initial data calibration was performed by ALMA staff using standard procedures in \texttt{CASA} (Common Astronomy Software Applications) version 6.4 \citep{mcmullin_casa_2007}. Additionally, to increase the signal-to-noise (SNR) of the observations, we further self-calibrated the data using the continuum. The first step was to create pseudo-continuum visibilities by flagging all channels that contained line emission. 
Then, we ran a total of seven phase-only calibration iterations and one amplitude calibration iteration. 
This procedure improved the SNR of the continuum emission only by a factor of $\sim 1.2$ for 216--0939 and $\sim 1.8$ for 253--1536A/B on average. Then, the self-calibration solutions were applied to each spectral window including the channels with line emission. 
The continuum was then subtracted from the visibilities using the \texttt{uvcontsub} routine to produce the self-calibrated visibilities of the different lines. 

\begin{figure*}[t!]
\centering
\includegraphics[width=16cm]{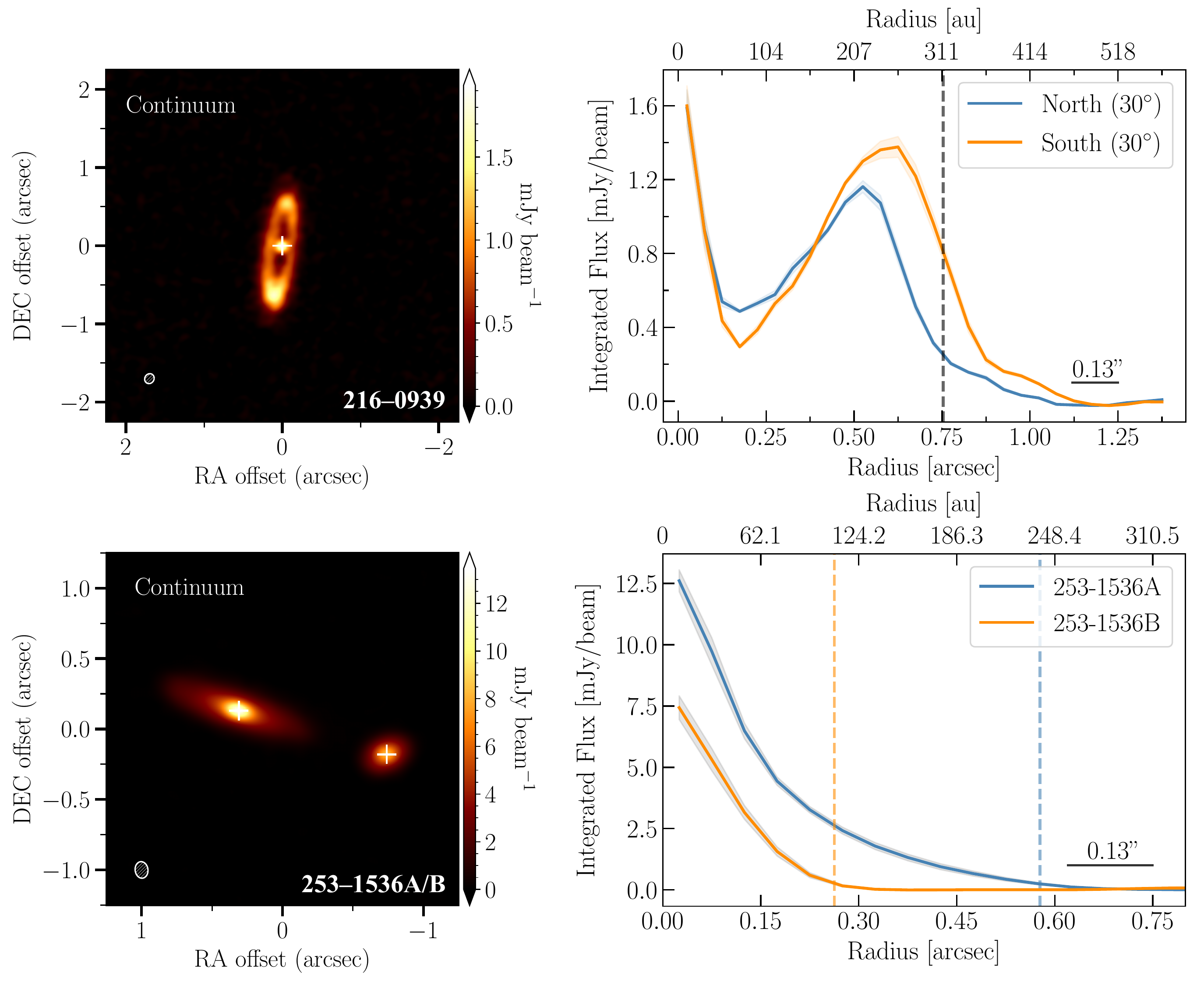}
\caption{{\em{Left:}} Dust continuum emission images at high-angular resolution from the protoplanetary disks 216--0939 (\textsl{top}) and 253--1536 (\textsl{bottom}). {\em{Right:}} Deprojected radial profiles for the continuum emission of each disk. 
For 216--0939, the radial profile is divided in north (blue) and south (orange) side emission. In particular, we estimated each radial profile including the emission from an aperture of $30^\circ$ in each side. For 253--1536, the radial profile of 253--1536A and 253--1536B are shown in blue and orange, respectively. Vertical lines represent the estimated disk size for each case, containing $95\%$ of the total flux of each disk. For 216--0939, the black vertical line represents the disk size estimated from the full azimuthally averaged radial profile.
\label{fig:continuum}} 
\end{figure*}

The continuum images were produced from the self-calibrated visibilities by first extracting the continuum channels using the \texttt{split} routine and imaging the visibilities using the \texttt{tclean} routine with a Briggs robust weighting parameter of $+0.5$, which balances the SNR and spatial resolution. We used an elliptical cleaning mask created using \texttt{CASA} regions for each source (216--0939, 253--1536A, and 253--1536B). The continuum images are shown in Fig.~\ref{fig:continuum}, and the final rms and beam sizes of the continuum images are listed in Table \ref{table:spectroscopic-beam}.

For the molecular lines, we used a Briggs robust parameter value of $+0.5$ because the line emission was not bright enough to be imaged at higher angular resolution. We also used the \texttt{uvtaper} parameter in \texttt{tclean} to reduce the weight of the longest baselines in the {$uv-$plane,} in particular we used \texttt{uvtaper=0.5 arcsec}. In this way, higher spatial frequencies are weighted down relative to lower spatial frequencies, increasing the sensitivity to larger-scale emission. This technique is used when there are poorly sampled areas in the $uv-$plane or to increase the SNR.
We also explored the effects of {$uv$} cuts to minimize the cloud contamination, which affected some of the lines. The same was investigated by \cite{factor_alma_2017} for the HCO$^+$ and HCN lines (4-3). They excluded baselines shorter than $70~k\lambda$ and found some improvement in the quality of the observations. However, we did not find a significant improvement in the quality of our observations after removing the shortest baselines. This can be explained by the different baseline coverage, which spans from 41 to 3640 m in our dataset, while it spans from 21.2 to 384.2~m in the data analyzed by \cite{factor_alma_2017}.

To help the cleaning process, we created a Keplerian mask adapted to each source in the \texttt{tclean} process task, that selects regions with line emission in each channel. 
The Keplerian masks were generated with the publicly available \texttt{Python} code from \citet{rich_teague_2020_code}, that computes the Keplerian motion of the disk given the mass of the central star, the disk geometry  and the cube parameters, such as spectral resolution, line frequency, source position, and systemic velocity. 
For the binary system, we created a Keplerian mask for 253--1536A and 253--1536B separately\footnote{For this we used a modified version of the code from \url{https://github.com/kevin-flaherty/ALMA-Disk-Code}} and then added them to produce a total mask that was used in the cleaning process. The inclination and PA used to create the masks are listed in Table~\ref{table:disk-properties}. However, the reported PA for 253-1536B disk did not capture well the Keplerian rotation of the disk, so after some visual inspection we instead used a more conservative mask with an inclination of $45^{\circ}$ and PA of $-90^\circ$. Appendix \ref{sec:appendixB} shows the resulting Keplerian masks overlaid on the channel maps. The rms and beam sizes of the line cubes can be found in Table \ref{table:spectroscopic-beam}.

In order to better compare the dust continuum emission with the molecular line emission, we created a second version of the dust continuum images, using the \texttt{imsmooth} routine to degrade the angular resolution of the continuum observations. This task performs a Fourier-based convolution to smooth the image and increase the SNR. In particular, we used it to obtain dust continuum images with the same angular resolution as the images of the different molecular lines.

\section{Results} \label{sec:results}
In this section, we present the continuum and molecular line detections and non-detections in the two externally irradiated systems. First, we describe the high-resolution continuum emission of the disks. Then we present the results for the molecular line emission. In particular, we extract disk-integrated flux densities for the molecular lines, and disk-averaged column densities. Finally, we estimate the HCN excitation temperature for both disks, using archival Band 7 observations.

\subsection{Dust continuum emission}
\begin{figure*}[ht]
\centering
\includegraphics[width=18cm]{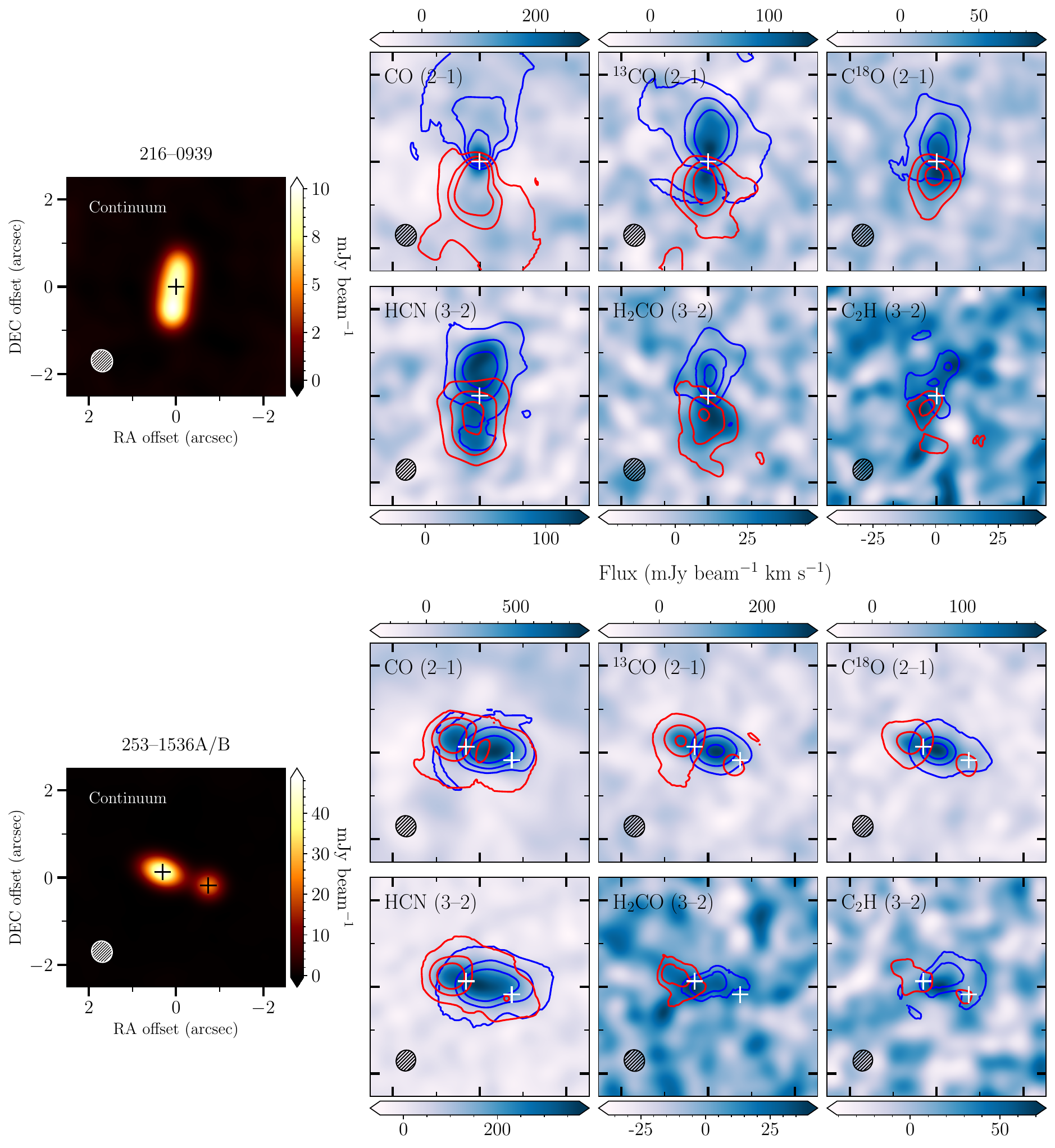}
\caption{
{\em{Left:}} Smoothed dust continuum emission for the 216--0939 \textsl{(top)} and 253--1536A/B \textsl{(bottom)} disks.
{\em{Right:}} Moment zero maps for CO $(2-1)$, $^{13}$CO $(2-1)$, C$^{18}$O $(2-1)$, HCN~$(3-2)$, H$_2$CO $(3-2)$, and C$_2$H~$(3-2)$, integrated over the full line width (colormap) and over the red- and blue-shifted parts of the line (contours). Labels and ticks are the same as those in the dust continuum images
For 216--0939, CO $(2-1)$ contour levels correspond to 3, 18, and 32$\sigma$, and for $^{13}$CO $(2-1)$ and C$^{18}$O $(2-1)$ contours correspond to 3, 12, and 24$\sigma$. On the other hand, HCN~$(3-2)$ contour levels correspond to 3, 12, and 24$\sigma$, for H$_2$CO $(3-2)$ they correspond to 3, 9, and 15$\sigma$, and for C$_2$H~$(3-2)$ contours correspond to 3 and 6$\sigma$.
For 253--1536, CO $(2-1)$ contours correspond to 3, 48, and 120$\sigma$, and for $^{13}$CO $(2-1)$ and C$^{18}$O $(2-1)$ contours correspond to 3, 28, and 48$\sigma$. Finally, HCN~$(3-2)$ contour levels correspond to 3, 18, and 40$\sigma$, and for H$_2$CO $(3-2)$ and C$_2$H~$(3-2)$ contours correspond to 3 and 6$\sigma$.
\label{fig:moment_maps_COisotopologues}} 
\end{figure*}

The beam size of the high-angular resolution continuum images of our sources is $\sim0\farcs13$, resulting in a spatial resolution of $\sim54$~au (for a distance of 414~pc). These images are shown in the left panels of Fig.~\ref{fig:continuum}. The smoothed continuum images have an angular resolution of $\sim0\farcs$5 and are shown in the left panels in Figure \ref{fig:moment_maps_COisotopologues}. 

The new high-angular resolution observations allow us to spatially resolve the dust emission in these disks. Using the Python package \texttt{GoFish} \citep{teague_gofish_2019} to generate deprojected radial profiles (see right panels in Fig.~\ref{fig:continuum}), we estimate a disk size of $311.5\pm 14.5~\mathrm{au}$ for the 216--0939 disk, defined as the radius containing $95\%$ of the total flux. The observations also reveal a central cavity with an outer edge at $120-135$~au. The cavity is also clearly seen in the cleaned image (see Figure \ref{fig:continuum}). We note that this inner cavity is consistent with the scenario of the central star being a tight binary system, that would clear the inner disk from material due to tidal interactions with the disk, which was proposed by \citet{factor_alma_2017}. Additionally, the 216--0939 disk seems to be asymmetric or eccentric, with the southern side being more elongated and $23\%$ brighter than the northern side. The origin of this eccentricity is unknown and should be investigated in the future. 

We are able to spatially resolve the region between the two members of the binary system 253--1536A/B. We estimate an angular separation of 0\farcs3 between the two disk edges (equivalent to $\sim124$~au), and a disk size of $239.1\pm14.5$~au and $108.7\pm14.5$~au for 253--1536A and 253--1536B, respectively. At the current angular resolution, the continuum emission looks very symmetrical for both disks, contrary to other binary systems where spirals have been observed due to the interaction between the disks \citep[e.g.,][]{kurtovic_disk_substructures_2018}. However, substructure could appear with even higher angular resolution observations.

\subsection{Molecular lines}\label{sec:molecular-lines}
In this section, we present the results for the observed molecular lines in both disks. First, we explain our criteria to determine whether a line is detected or not. We then describe the spatial distribution of the detected lines. 

\subsubsection{Line detections}\label{subsubsec:line-derection}

A line was considered to be detected if emission within the Keplerian mask was $\ge 3\sigma$ (rms) in at least $3$ velocity channels. Following this criteria, $^{12}$CO $(2-1)$, $^{13}$CO $(2-1)$, C$^{18}$O $(2-1)$, and HCN~$(3-2)$ are robustly detected in both sources. Indeed, $>3\sigma$ emission is detected in almost every channel in both sources, with $10\sigma$ emission detected in at least 3 channels. H$_{2}$CO $(3-2)$ and C$_2$H~$(3-2)$ are also detected according to this criteria but at lower SNR. In particular, $5\sigma$ emission is detected in a couple of channels for H$_{2}$CO $(3-2)$, and C$_2$H $(3-2)$ is only detected at the $3\sigma$ level in five or six channels in both disks. The other observed lines did not show emission $\ge 3\sigma$ in any channel, and are therefore considered as non-detections. For examples of channel maps of detected and non-detected lines, see Appendix \ref{sec:appendixB}.

To confirm the detection of H$_2$CO and C$_2$H, we used the matched-filter method described in \citet{loomis_detecting_2018}. The advantage of this method is that it looks for a signal directly in the visibilities, and therefore removes the uncertainties of the cleaning process which is present in the images. This method requires a model image as a matched filter; we used the HCN~$(3-2)$ and C$^{18}$O $(2-1)$ cubes as filters, as these molecular lines were robustly detected and they are less affected by the cloud contamination. A molecular line is considered detected if a $\ge3\sigma$ peak is found near the source velocity in the filter-response spectrum, using at least one of the filters\footnote{We note that the filter response spectrum depends on the used filter because the method assumes that the two lines (weak data and filter) have the same spatial distribution. If this is not the case, it is possible to obtain a false negative (see Fig.~\ref{fig:visible_C2H} for an example).}. Following this criteria, we confirm the detection of both H$_2$CO and C$_2$H $(3-2)$ lines in both disks. We applied the same method to the lines that were not detected in the image plane, and confirm that DCN $(3-2)$, $c-$C$_3$H$_2$ $(6_{06} - 5_{15})$, $c-$C$_3$H$_2$ $(6_{25} - 5_{14})$, $c-$C$_3$H$_2$ $(7_{07} - 6_{16})$ and N$_2$D$^+$ $(3-2)$ are not detected. Fig.~\ref{fig:visible_H2CO_DCN} shows an example of the filter response for H$_2$CO $(3-2)$ (detected) and DCN $(3-2)$ (not detected) using HCN $(3-2)$ line as the filter.

\begin{figure*}[t!]
\centering
\includegraphics[width=0.8\textwidth]{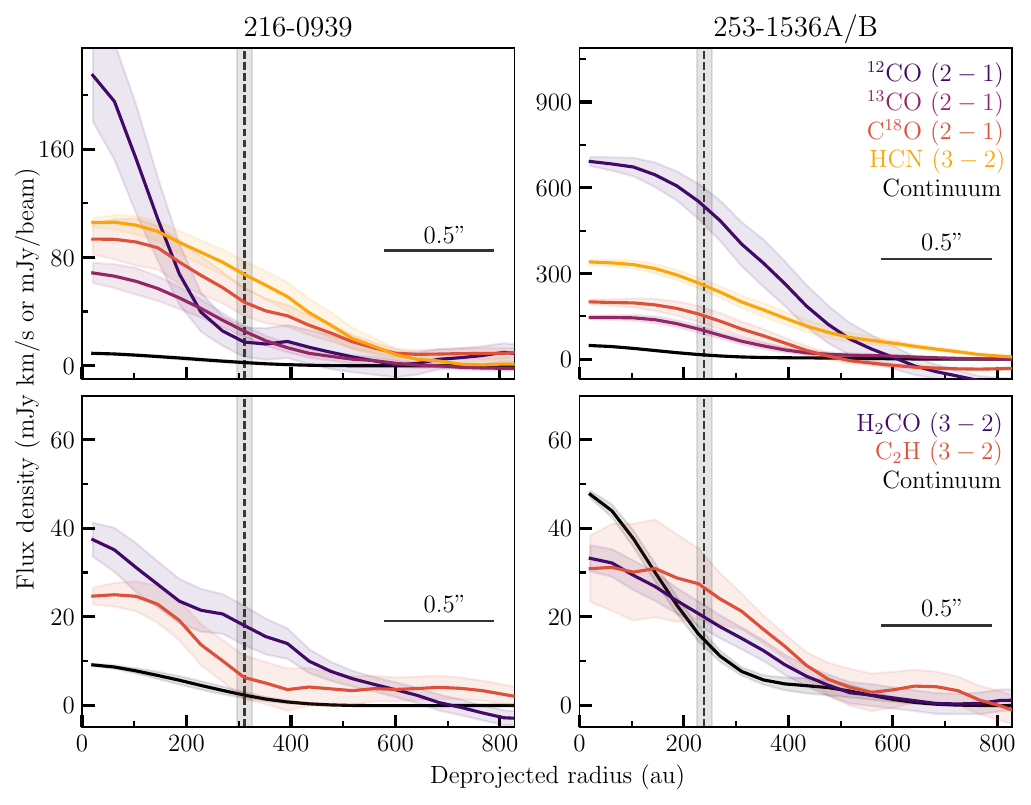}
\caption{
Deprojected radial and azimuthally averaged profiles for 216--0939 \textsl{(left)} and 253--1536A/B \textsl{(right)}.
\textsl{Upper panels:} $\mathrm{^{12}CO~(2-1)}$, $\mathrm{^{13}CO~(2-1)}$ and $\mathrm{C^{18}O~(2-1)}$, $\mathrm{HCN~(3-2)}$. \textsl{Lower panels:} $\mathrm{H_2CO~(3-2)}$ and $\mathrm{C_2H~(3-2)}$. The profile of the smoothed dust continuum emission is shown in black in all panels, in units of mJy/beam. Color-shaded regions show the 1$\sigma$ scatter of the averaged emission at each radial bin. The black vertical line marks the edge of the high-resolution dust continuum emission on each system, defined as the radius within which the $95\%$ of the total flux arises. In the case of the binary system, the radial profiles are obtained with 253--1536A as the center of the emission, and the dust edge shown corresponds to 253--1536A.
\label{fig:radial_profiles}} 
\end{figure*}

\subsubsection{Spatial distribution}

We created velocity-integrated maps for each detected molecular line. Figure~\ref{fig:moment_maps_COisotopologues} shows the resulting maps for 216--0939 (top) and 253--1536A/B (bottom). In this figure, the low-angular resolution images of the continuum emission are shown in the left panels for comparison. The top panels show the $^{12}$CO $(2-1)$, $^{13}$CO $(2-1)$, and C$^{18}$O $(2-1)$ line emission, and the bottom panels show the emission from HCN~$(3-2)$, H$_2$CO $(3-2)$ and C$_2$H~$(3-2)$. The red and blue contours show the Keplerian rotation of the disks, created by integrating the line emission over the red- and blue-shifted parts of the line. To do this, we integrated the emission from the first channel that presented $>3\sigma$ line emission to the channel that was closest to the systemic velocity of the source (blue-shifted); the same was done to the red-shifted part, integrating the channels from the systemic velocity to the last channel that presented $>3\sigma$ line emission. We find that the $^{12}$CO line emission is more extended than the $^{13}$CO and C$^{18}$O line emission in both 216--0939 and 253--1536A/B systems. Moreover, the CO isotopologue emission is more extended compared to the dust continuum emission. The line emission from H$_2$CO and C$_2$H is weaker than the HCN and CO isotopologue line emission for both systems. In addition, we note that the $^{12}$CO line emission is heavily contaminated by extended cloud emission and foreground absorption near the systemic velocity. 
To a lesser degree, the other CO isotopologues and HCN also suffer from cloud contamination (see the channel maps in Appendix \ref{sec:appendixB} and the disk-integrated spectra in Appendix~\ref{sec:app_spectra}).

In the velocity-integrated maps it is also possible to disentangle the Keplerian rotation of the smaller disk from the larger disk in 253--1536A/B, in particular for the CO isotopologues and HCN, where the emission of the minor companion can be observed in the red- and blue-shifted contours (see also the channel maps in Appendix~\ref{sec:appendixB}). The emission can also be tentatively disentangled for C$_2$H. Figure \ref{fig:CO_mom1} shows the first moment map for the CO $(2-1)$ line in 253--1536A/B, where a tentative deviation from Keplerian rotation can be seen towards the northern side of 251--1536A (black arrow). This deviation could be associated with the interaction between the two companions or be due to the cloud contamination. Unfortunately, the angular resolution of 0\farcs5 is not high enough to resolve the separation between the two disks and to disentangle if smaller spatial signals of dynamical interaction are present.

To further investigate the spatial distribution of the different molecular lines, we generated radial profiles using \texttt{GoFish} to deproject and azimuthally average the line emission in the zero moment maps, using the disk parameters. Figure \ref{fig:radial_profiles} shows the radial profiles of the brighter (top panels) and weaker (bottom panels) lines. For both 216--0939 and 253--1536A/B disks, all molecular line emission is more extended than the dust continuum emission. Most of the lines show little sub-structure at the current angular resolution (the apparent gaps/rings in C$_2$H~$(3-2)$ are probably related to the noise in the images). However, an interesting feature is that the central emission of the C$_2$H line is flat (and not centrally peaked), suggesting the emission could be arise from a ring. C$_2$H ringed emission has been observed in other isolated disks \citep[e.g.,][]{bergin_hydrocarbon_2016,cleeves_tw_2021, pegues_alma_2021, guzman_molecules_2021}.

\begin{deluxetable*}{@{\extracolsep{10pt}}cccccccc}[t!]
\tablecaption{Disk integrated fluxes.}
\tablewidth{0pt}
\tablehead{
& & \multicolumn{1}{c}{\bf{216--0939}} & \multicolumn{1}{c}{\bf{253--1536A/B}}\\
\colhead{Molecule}   & \multicolumn{1}{c}{{Line}}  & \multicolumn{2}{c}{Integrated Intensity}    \\
                    &                       & \multicolumn{2}{c}{[mJy km s$^{-1}$]}
}
\startdata
\hline
\multicolumn{4}{c}{{Detected}} \\
\hline
\hline
$^{12}$CO   & $2-1$                 & $1261 \pm 218$       & $3381 \pm 1163$    \\
$^{13}$CO   & $2-1$                 & $1020 \pm 485$       & $610 \pm 567$      \\  
C$^{18}$O   & $2-1$                 & $341 \pm 144$        & $543 \pm 97$      \\
C$_2$H      & $N = 3-2, J = \frac{5}{2}-\frac{3}{2}, F = 3-2$                 & $159 \pm 76$         & $211 \pm 87$      \\
HCN         & $3-2$                 & $989 \pm 145$       & $2193 \pm 255$     \\
            & $4-3$\tablenotemark{b}                 & $1276 \pm 212$       & $3099 \pm 338$     \\
H$_2$CO     & $3_{22} - 2_{21}$    & $283 \pm 76$         & $106 \pm 54$       \\
\hline
\multicolumn{4}{c}{Non--detected\tablenotemark{a}} \\
\hline
\hline
$c-$C$_3$H$_2$  & $6_{06}-5_{15}$    & $< 97$   & $< 107$   \\
            & $6_{25}-5_{14}$    & $< 138$   & $< 183$   \\
            & $7_{07}-6_{16}$    & $< 126$   & $< 132$   \\
DCN         & $3-2$                 & $< 112$   & $< 117$   \\
N$_2$D$^+$  & $3-2$                 & $< 79$    & $< 169$   \\
\hline
\enddata
\tablecomments{The integrated fluxes are measured within the Keplerian masks. In the case of 253--1536, they include both A and B members.
}
\tablenotetext{a}{Reported fluxes for non-detections correspond to $3\sigma$ upper limits where $\sigma$ is the uncertainty estimated via bootstrapping.}
\tablenotetext{b}{The same process was performed on the ALMA Band 7 data obtained from Cycle 0 project \#2011.0.00028.s \citep{mann_alma_2014}}
\label{table:spectroscopic-intensities}
\end{deluxetable*}

\subsubsection{Disk integrated fluxes}

Disk-integrated fluxes were estimated for the detected lines, and upper limits are reported for non-detections. To calculate the integrated flux of each line, we first multiplied the image cubes by the same Keplerian masks used to clean the data, using the \texttt{immath} routine in \texttt{CASA}, and then summed over the line emission. This additional step removes some of the noise in the image and results in a better-integrated disk spectrum where the Keplerian rotation is shown. To estimate the uncertainty in the integrated flux for each line, we used the same method described above to extract the fluxes but now changing the disk center in $1000$ random samples in regions without emission outside of the original Keplerian mask via bootstrapping. The uncertainty in the integrated flux was then computed as the standard deviation of the resulting distribution added in quadrature with a $10\%$ systematic flux calibration uncertainty. The number of samples is chosen to be sufficiently large so that the estimated uncertainty does not vary significantly. The resulting integrated intensities of the detected lines are listed in Table \ref{table:spectroscopic-intensities}. We also list upper limits for the non-detected lines. The CO and CO isotopologue fluxes have larger errors because of the cloud contamination. Indeed, the cloud is more pronounced in CO than in the other species.

\subsection{Column density retrieval}

To estimate the disk-averaged column densities ($N_\mathrm{T}$) we consider that the gas obeys local thermal equilibrium (LTE), and assume a range of possible excitation temperatures ($T_\mathrm{ex}$) based on typical gas temperatures observed in the outer regions of disks \citep{guzman_molecules_2021}. 
The LTE assumption is reasonable because typical gas densities in disks are high compared to the critical densities of the observed HCN, H$_2$CO, and C$_2$H lines\footnote{For gas temperatures between $10$ and $50$ K, the HCN~$(3-2)$, H$_2$CO $(3-2)$, and C$_2$H~$(3-2)$ lines, have a critical density of $\sim7\times 10^7$, $(2-3)\times 10^6$, and $(5-6)\times 10^6~\mathrm{cm^{-3}}$, respectively.}, assuming their emission arises mainly from layers that are close to the midplane \citep{law_molecules_2021b}.

Under LTE conditions, the energy levels are populated following Boltzmann's law, 

\begin{equation}
N_u = \frac{N_\mathrm{tot}}{Q(T_\mathrm{ex})} g_u \exp{\left(\frac{-E_u}{k_\mathrm{B}T_\mathrm{ex}}\right)},
\label{eq:rot-diagram}
\end{equation}

\noindent where $N_u$ corresponds to the upper-level column density, $N_\mathrm{tot}$ is the total column density of the molecule, $Q(T_\mathrm{ex})$ is the partition function, $g_u$ is the upper-level degeneracy, $E_u$ is the upper-level energy, and $k_\mathrm{B}$ is the Boltzmann constant. If the line is optically thin, the upper-level column density can be written as

\begin{equation}
N_u^\mathrm{thin} = \frac{8\pi k_\mathrm{B}\nu^2 W}{hc^3A_{ul}},
\label{eq:nuthin}
\end{equation}

\begin{figure*}[t!]
\centering
\includegraphics[width=0.6\textwidth]{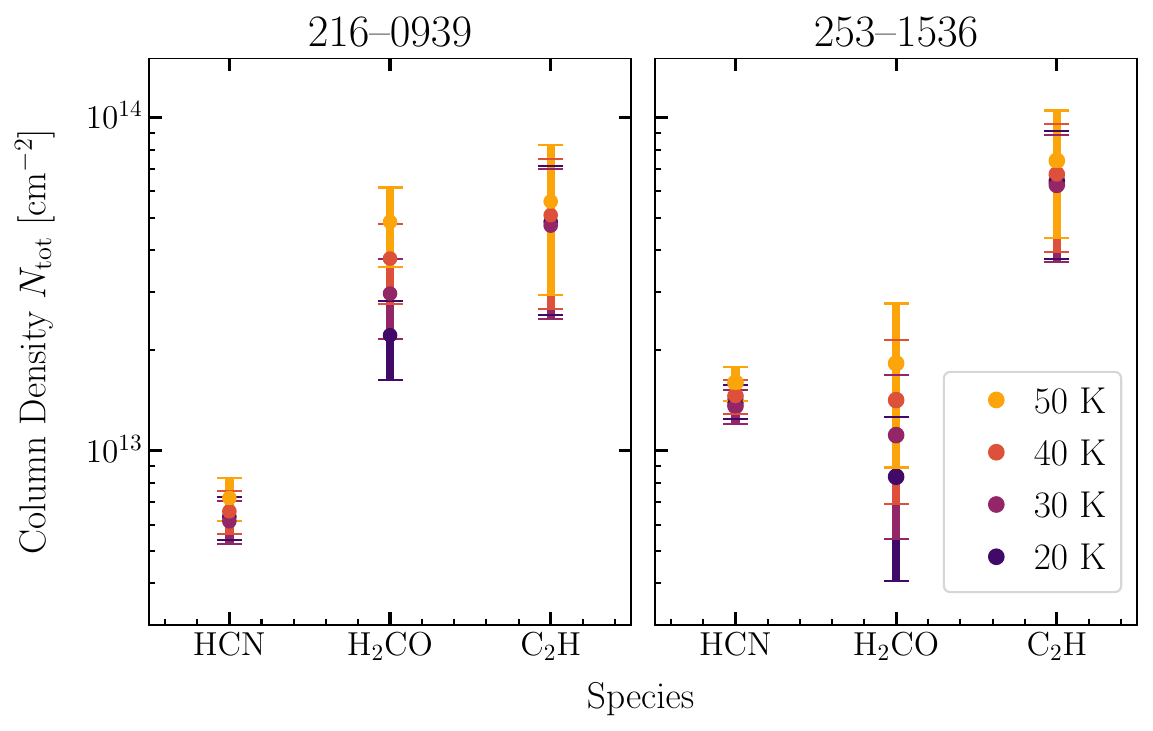}
\caption{Disk-averaged column densities for HCN, H$_2$CO and C$_2$H, assuming optically thin emission. The colors represent different excitation temperatures ($T_\mathrm{ex}$) assumed in the calculation.}
\label{fig:column_density} 
\end{figure*}

\noindent where $\nu$ is the line frequency, $W$ is the integrated line intensity, $h$ is the Planck constant, $c$ is the speed of light, and $A_{ul}$ is the Einstein coefficient for spontaneous emission. All frequencies, Einstein coefficients, and partition functions values were taken from the CDMS catalog \citep{muller_cologne_2001, muller_cologne_2005, endres_cologne_2016}, and can be found in Table \ref{table:splatalogue}.

We assume the optically thin approximation and a disk-averaged excitation temperature that ranges from $20-50$ K, which are typical temperatures found in the disk molecular layer \citep[e.g.,][]{walsh_chemical_2010}. The results are shown in Fig.~\ref{fig:column_density}. We note that for the \mbox{253--1536A/B} system, the reported column densities include the contribution from both members A and B. We can see that the HCN, H$_2$CO, and C$_2$H column densities are not too sensitive to the excitation temperature assumptions in this range. \textcolor{black}{Overall, we find 
$N_\mathrm{tot}(\mathrm{HCN}) \sim (0.5-1.8) \times 10^{13}$~cm$^{-2}$, 
$N_\mathrm{tot}(\mathrm{H}_2\mathrm{CO}) \sim (0.4-6.2) \times 10^{13}$ cm$^{-2}$, and 
$N_\mathrm{tot}(\mathrm{C}_2\mathrm{H}) \sim (0.3 - 1.0)\times 10^{14}$ cm$^{-2}$ 
for both disks}. These column densities have to be considered as lower limits because we are assuming that the lines are optically thin, which might not be the case, in particular for HCN.

We find that the HCN column density is quite similar between the two disks, with differences of only a factor of $\sim$2. Moreover, the C$_2$H column density is almost the same for the two systems. In contrast, the formaldehyde column density varies between the disks. In particular, the H$_2$CO/HCN ratio is a factor $\sim 5$ larger in 216--0939 compared to 253--1536A/B. These column densities are within the range of what has been found in previous studies of isolated disks \citep[e.g.,][]{pegues_alma_2020, guzman_molecules_2021}, which reported disk integrated column densities for HCN and C$_2$H in the range of $10^{12}-10^{15}$ cm$^{-2}$ in disks around low-mass stars, and $>10^{16}$ cm$^{-2}$ in Herbig Ae/Be disks. 

\subsection{HCN excitation temperature}

\begin{figure}[b!]
\centering
\includegraphics[width=0.5\textwidth]{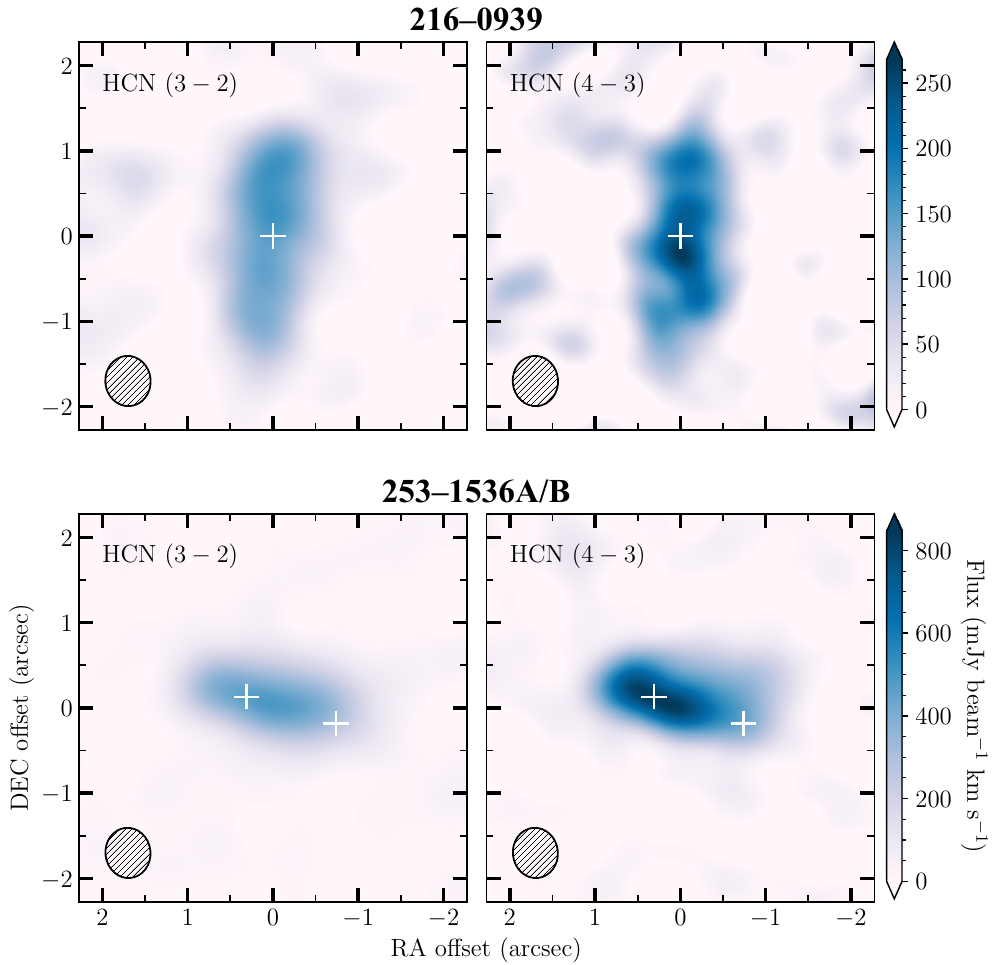}
\caption{Zeroth--moment maps of HCN~$(3-2)$ \textsl{(left)} and HCN $(4-3)$ \textsl{(right)} line emission of the two disks. \textsl{Top:} 216--0939, \textsl{Bottom:} 253--1536A/B.}
\label{fig:rotational_diagram_mol}
\end{figure}

\begin{figure*}[t!]
\centering
\includegraphics[width=0.8\textwidth]{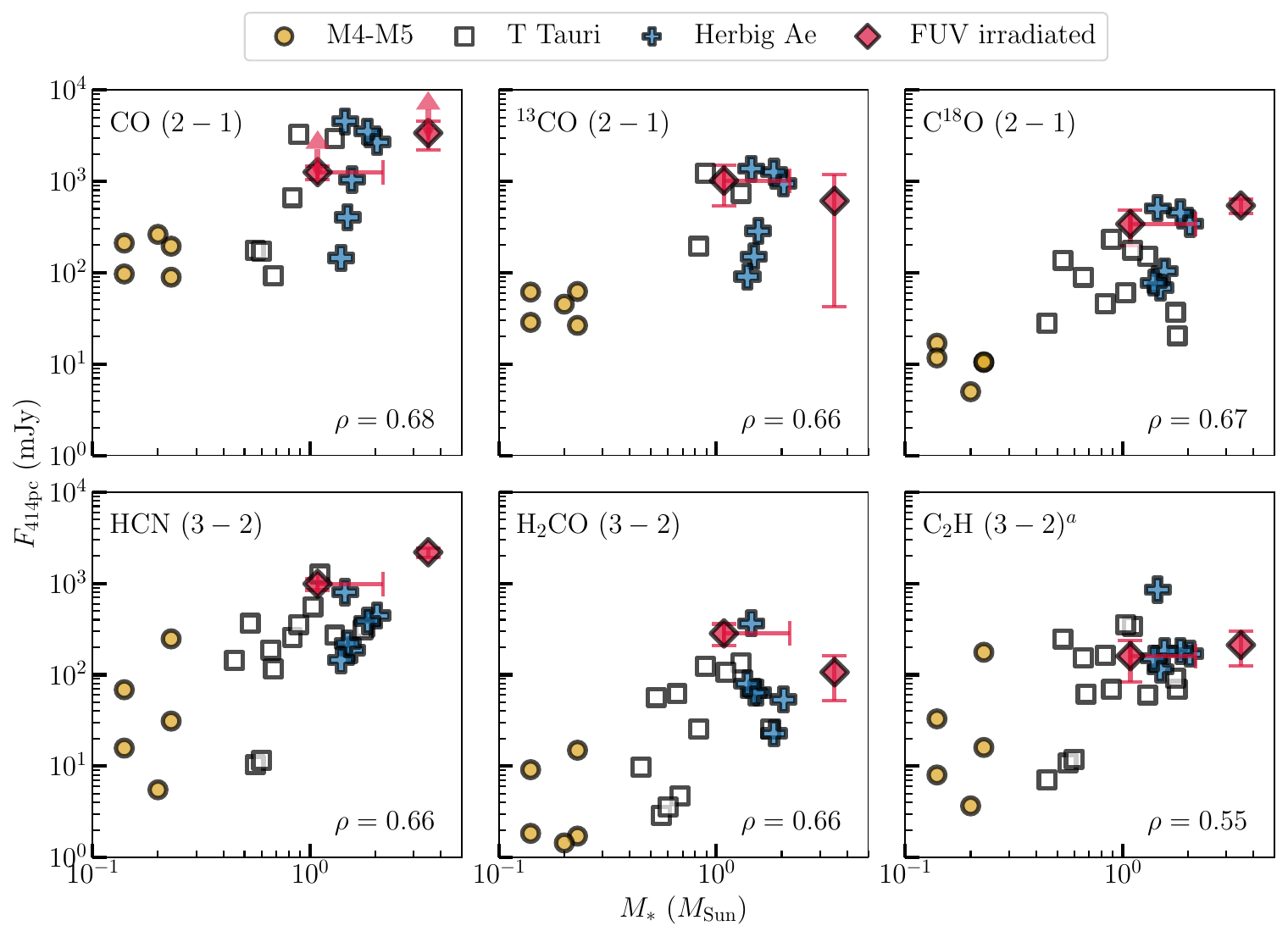}
\caption{
Molecular line fluxes for FUV irradiated disks in the Orion Nebula Cluster (this work) as a function of stellar mass, compared with isolated disks compiled from the literature, including M4-M5 stars \citep{pegues_alma_2021}, T Tauri and Herbig Ae/Be \citep{huang_alma_2017, bergner_organic_2019, bergner_survey_2019, bergner_evolutionary_2020, pegues_alma_2020, oberg_molecules_2021, law_molecules_2021, guzman_molecules_2021, pegues_sma_2023}. The fluxes have been normalized to a common distance of $414$ pc. 
The left red diamond corresponds to the 216--0939 disk, assuming the source is a binary system of two $\sim1~\mathrm{M_\odot}$ stars. A horizontal error bar is added to consider the possibility the system is a single massive star of $2.17~\mathrm{M_\odot}$. The right red diamond corresponds to the 253--1536A disk. For CO~$(2-1)$, lower limits are shown since these lines are highly affected by cloud contamination. 
$^a$ C$_2$H~$(3-2)$ includes $J=5/2-3/2$ and $J=7/2-5/2$ lines.
\label{fig:fluxes}} 
\end{figure*}

We estimated the excitation temperature of HCN using the ratio between the HCN~$(3-2)$ observations presented in this work and the HCN $(4-3)$ line initially published in \cite{mann_alma_2014} and later presented in more detail in \cite{williams_alma_2014} and \cite{factor_alma_2017}. Figure \ref{fig:rotational_diagram_mol} shows the HCN~$(3-2)$ and $(4-3)$ moment zero maps for both sources. For consistency, we computed the disk-integrated flux of the HCN $(4-3)$ line in the same manner as was done for the $(3-2)$ line. Additionally, we smoothed the HCN~$(3-2)$ emission to match the angular resolution of the HCN $(4-3)$ line $(0\farcs59\times 0\farcs53)$. We used the spectroscopic parameters of HCN listed in Table \ref{table:splatalogue}, and assumed $\tau=0$ (the optically thin case).
We infer an excitation temperature of $20.0^{+4.1}_{-2.9}$~K and $22.5^{+5.4}_{-3.6}$~K for 216--0939 and 253--1536A/B, respectively. The derived excitation temperature of $\sim20$~K is similar for both disks and also similar to what has been found for other disks \citep[e.g.,][]{bergner_survey_2019, guzman_molecules_2021}, suggests that the HCN emission arises from a relatively cold layer close to the midplane. 

In order to investigate whether the excitation temperature increases in the outer disk, which could be expected for externally irradiated disks, and taking advantage of the angular resolution of the observations, we also derived the excitation temperature as a function of radius. We found that, at the current angular resolution, the excitation temperature is constant across both disks, with no visible increase in the outer disk.

\section{Discussion} \label{sec:discussion}
In this section, we first compare the measured line fluxes of 216-0939 and 253-1536A/B with those found in disks around stars of different masses that are located in low-mass star forming regions, where the external radiation field is significantly lower compared to the disks in our sample. Then, we compare our findings with predictions from chemical models.

\subsection{Comparison between irradiated and isolated disks}

\begin{figure}[t!]
\centering
\includegraphics[width=0.5\textwidth]{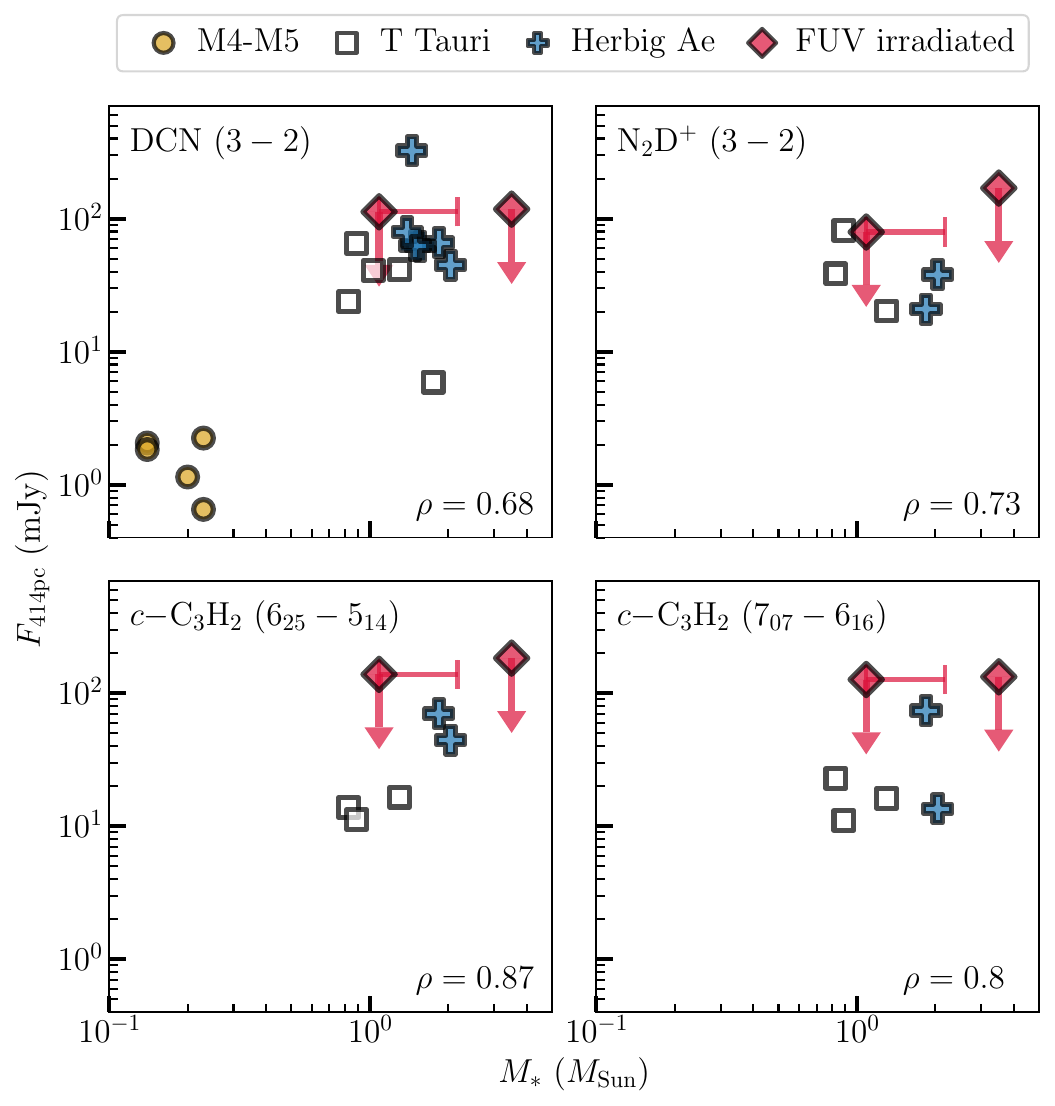}
\caption{
Same as Figure \ref{fig:fluxes}, but displaying the upper limits for a subset of the non-detected molecular lines. }
\label{fig:unfluxes}
\end{figure}

Figures \ref{fig:fluxes} and \ref{fig:unfluxes} show the distance-normalized integrated fluxes and upper limits as a function of stellar mass, for the detected and non-detected lines, respectively. 
For 216--0939, we consider two possible stellar scenarios: a tight equal mass binary of two 1.1~M$_\odot$ stars and a single star of $2.17$~M$_\odot$, as discussed in Section \ref{sec:obs}. For 253--1536A/B, we only consider the mass of the primary A star ($3.5$~M$_\odot$), since most of the emission we detect comes from this source. 
For comparison, we also include fluxes reported for isolated disks around M4--M5 stars, T Tauri stars, and Herbig Ae/Be stars. The literature disk sample was compiled from \cite{huang_alma_2017}, \cite{bergner_organic_2019}, \cite{bergner_survey_2019}, \cite{bergner_evolutionary_2020}, \cite{pegues_alma_2020}, \cite{pegues_alma_2021}, \cite{law_molecules_2021}, \cite{guzman_molecules_2021}, \citet{oberg_molecules_2021}, and \cite{pegues_sma_2023}. We note that the different  molecular line fluxes were compiled from different studies, and in some cases the same line flux is reported in two or more studies. In those cases, we selected the most recent study. The total disk sample includes $5$ M4--M5 stars, $6$ Herbig Ae disks, $13$ T Tauri disks, and the $2$ externally irradiated systems from our sample. The stellar masses range from $0.12$ to $3.5~\mathrm{M}_\odot$, and the stellar ages range between 0.4 to $>10$~Myr. The two disks in Orion have ages within this range \citep[$\sim1-3$~Myr;][]{williams_alma_2014,factor_alma_2017}.

Considering the combined sample of isolated and externally irradiated disks, we find a positive correlation between the fluxes and stellar mass, for all the lines presented here (see Fig.~\ref{fig:fluxes}). This trend was previously reported for isolated disks only by \cite{pegues_alma_2021} and \cite{pegues_sma_2023}. The estimated Spearman correlation coefficient of the combined disk sample is shown in the bottom right corner of each panel, which corresponds to the dispersion of the data and measures the correlation between the two variables.
The closer to one the value of the Spearman correlation coefficient, the stronger the correlation. The strongest correlations are found for C$^{18}$O, HCN and C$_2$H. In principle, one could have expected to find brighter CO emission in the two irradiated disks compared to the isolated disks because they are expected to be warmer. However, we find that the $^{12}$CO and $^{13}$CO lines are weaker than expected according to the observed trend, with an emission flux similar to the Herbig disks of lower masses. 
This is probably related to the cloud contamination in the irradiated disks, which results in a drop of the CO emission in the central channels (see the channels maps in Appendix~\ref{sec:appendixB}). This is also consistent with the fact that the less contaminated C$^{18}$O emission shows a stronger correlation than the more abundant CO isotopologues. This suggests that there is no difference in the chemistry between the isolated sources and the two irradiated disks presented here, and that the differences between the line fluxes in the disks are mainly due to the stellar masses.

Regarding the non-detected molecular lines, 
the derived upper limits are consistent with the fluxes measured in other disks, where these lines have been detected (see Figure \ref{fig:unfluxes}).

\begin{figure*}[t!]
\centering
\includegraphics[width=0.9\textwidth]{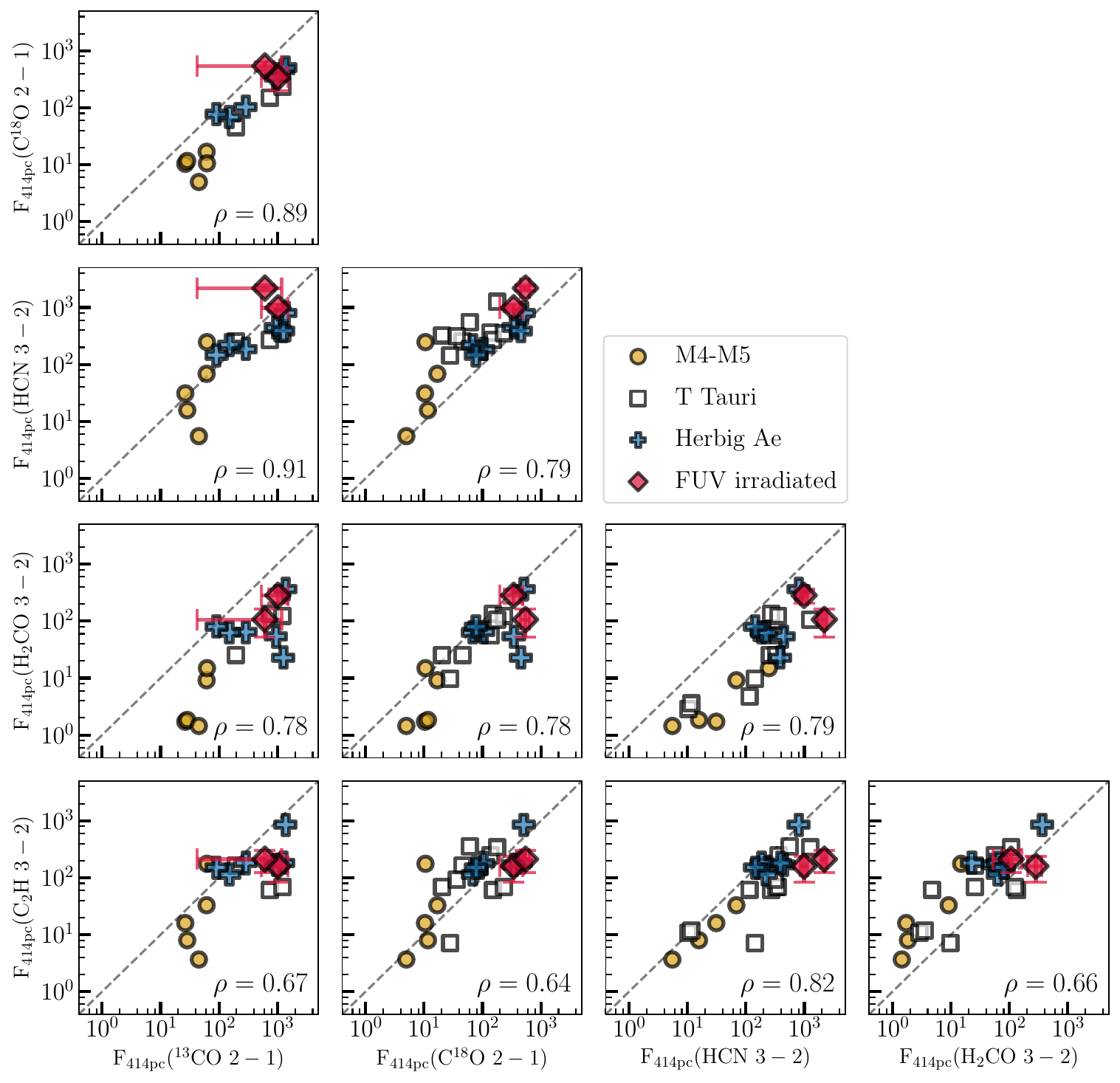}
\caption{The molecular line ratios for FUV irradiated disks in the Orion Nebula Cluster (this work) compared with isolated disks from the literature. The two disk systems from this work are shown as red diamonds. The dashed line shows a one-to-one correlation. $^{12}$CO is not included in this comparison because the derived fluxes are affected by cloud contamination.
\label{fig:ratios}}
\end{figure*}

In order to remove the dependence of the fluxes on the stellar mass, in Fig.~\ref{fig:ratios} we show the normalized line fluxes of each line against one another, in a similar manner to \cite{bergner_organic_2019}, \cite{bergner_survey_2019}, \cite{pegues_alma_2020}, \cite{pegues_alma_2021}, and \cite{pegues_sma_2023}.
Overall, we find that for the combined disk sample, every molecular line combination in Fig.~\ref{fig:ratios} has a strong and positive correlation, as was previously found by \citet{bergner_survey_2019, pegues_alma_2021,pegues_sma_2023} for the isolated disks only. The C$_2$H vs H$_2$CO pair shows one of the worst correlation, with a Spearman coefficient of $0.66$. Indeed, the observed C$_2$H flux in the irradiated disks is lower than expected based on the observed trend in isolated disks. The strongest correlation corresponds to HCN vs $^{13}\mathrm{CO}$ with a Spearman coefficient ($\rho=0.91$); followed by C$^{18}$O vs $^{13}$CO ($\rho=0.89$), and C$_2$H vs HCN, with a correlation coefficient of $0.82$. 

\begin{figure}[ht]
\centering
\includegraphics[width=0.5\textwidth]{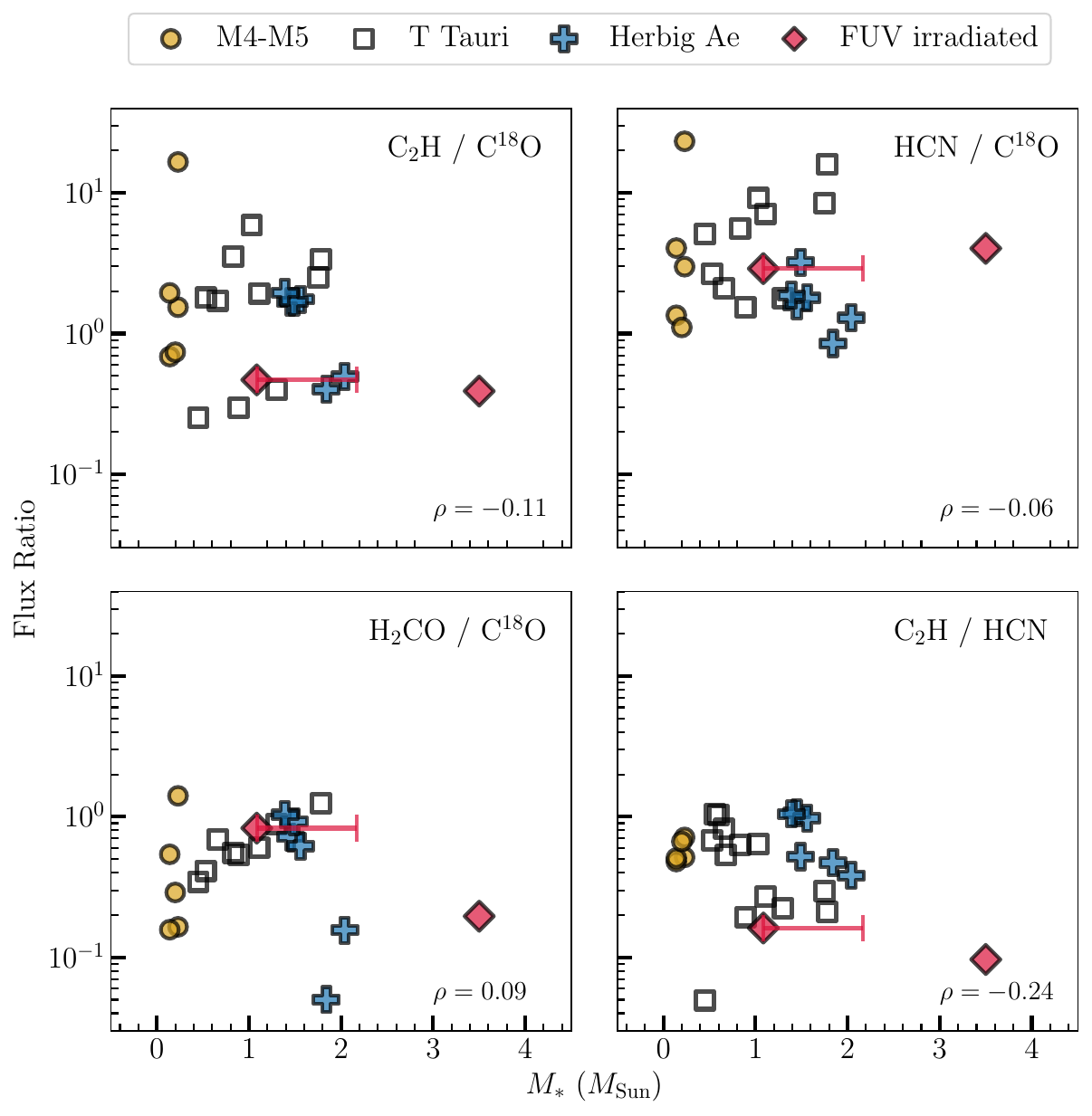}
\caption{The molecular line ratios for FUV irradiated disks in the Orion Nebula Cluster (this work) as a function of stellar mass compared with isolated disks from the literature. Note that this figure shows only a subset of the non detected lines.
\label{fig:ratiovsmass}}
\end{figure}

Finally, Figure \ref{fig:ratiovsmass} shows four different flux ratios (C$_2$H/C$^{18}$O, H$_2$CO/C$^{18}$O, HCN/C$^{18}$O, and C$_2$H/HCN) as a function of stellar mass.
Considering the isolated disks only, \cite{pegues_alma_2021} found no clear correlation between these flux ratios and stellar mass. However, with the addition of the externally irradiated disks 216--0939 and 253--1536A/B, we find that the C$_2$H~$(3-2)$/HCN~$(3-2)$ flux ratio shows a tentative trend, with the ratio decreasing with stellar mass (Spearman coefficient of $-0.24$). This result is unexpected since the C$_2$H emission should be brighter in irradiated disks because the C$_2$H formation is expected to be enhanced in the presence of FUV radiation. However, other factors, like carbon depletion \citep{bergin_hydrocarbon_2016} and dust growth/settling, also play a role in the formation of carbon chains, and could explain the faint C$_2$H emission in the two Orion disks. In addition, HCN is also known to be sensitive to photochemistry \citep{guzman_cyanide_2015, visser_nitrogen_2018, bergner_molecules_2021, pegues_alma_2021}. Future observations towards more irradiated disks are needed to confirm the observed tentative trend. 

\begin{figure*}[ht]
\centering
\includegraphics[width=16cm]{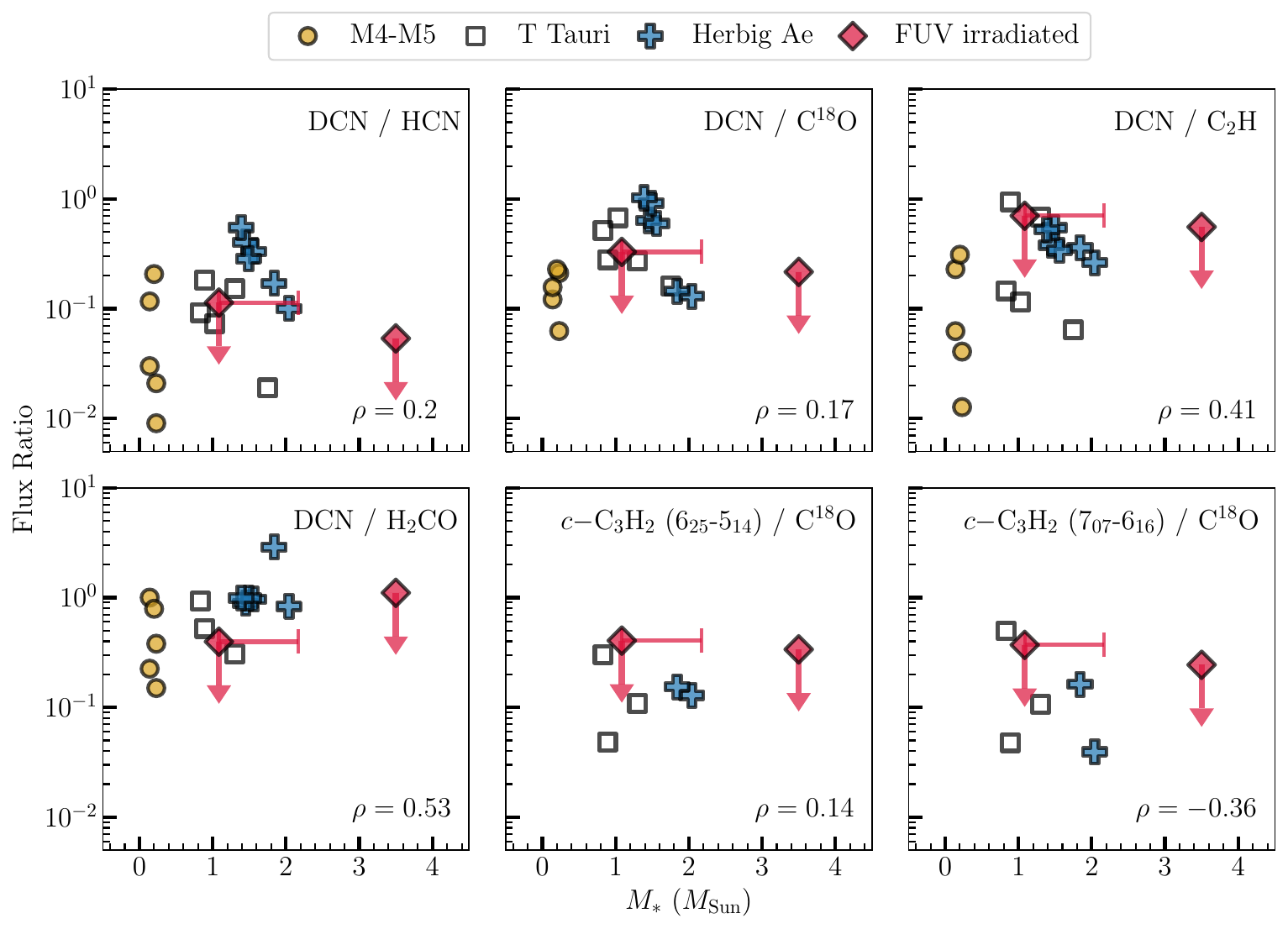}
\caption{Same as Figure \ref{fig:ratiovsmass} for a subset of the non--detected molecular lines. 
\label{fig:non_ratiovsmass}}
\end{figure*}

A similar comparison of flux ratios is presented in Figure \ref{fig:non_ratiovsmass} but for a subset of the non-detected molecular lines (DCN and $c-$C$_3$H$_2$). We find that the observed upper limits in the two irradiated disks are, in general, consistent with the values found in isolated disks. However, the ratios involving DCN (e.g., DCN/HCN and DCN/H$_2$CO) seem to be slightly lower for the irradiated disks compared to the isolated ones. If this is confirmed, it would suggest that irradiated sources have less cold material than isolated disks, which is consistent with the expectation of irradiated disks being warmer. 
However, in that case we would also expect to see an enhancement of H$_2$CO in the outer disk due to ice desorption in the warm gas, assumming that H$_2$CO was inherited, and this is not observed with the current observations (see Fig.~\ref{fig:radial_profiles}). Observations with better sensitivity are needed to confirm this result.

These results suggest that the chemistry of the two irradiated disk systems presented here is not too different from the chemistry of isolated sources, at least for the molecular lines considered here. Indeed, the two irradiated disks follow the trends observed for nearby isolated disks. However, we note that the results presented here may not be representative of the chemistry of all irradiated disks, because our sample consists of only two particularly massive irradiated disk systems; 216--0939 and 253--1536A/B are massive disks, located at $1.59$ and $0.92$ pc away from the radiation field source, respectively. 
Indeed, it is possible that some material in the outer disk has been stripped away by the external radiation field, and our observations are just tracing the part of the disk that has survived, and is no longer affected by the external radiation field. 
Another possibility is that the external radiation fields are lower than the ones assumed here, due to projection effects or to the extinction of UV photons by surrounding cluster material.

Molecular line surveys of smaller disks and disks that are closer to the massive stars are needed to draw more general conclusions about the chemistry of externally irradiated disks. Unfortunately, this is very challenging because lines are usually heavily contaminated by the emission from the molecular cloud. However, recent surveys have been able to detect CO and HCO$^+$ lines in disks close to the ONC (ranging between $0.03 - 0.15 ~\mathrm{pc}$), thanks to very sensitive and high angular resolution ALMA observations \citep{boyden_protoplanetary_2020}.

\subsection{Comparison to model predictions}

Our sample consists of two systems around intermediate-mass stars ($1-3.5$~$\mathrm{M}_\odot$); unfortunately, chemical models that include an external radiation field have not been developed for this type of star/disk systems yet. However, previous theoretical studies have found that the chemistry of T Tauri and Herbig disks are not too different \citep{agundez_chemistry_2018}. Therefore, we compare our results with models of disks around lower-mass T Tauri stars. We use the chemical models presented in \cite{walsh_molecular_2013} and \cite{walsh_complex_2014}, that were developed for a disk around a T Tauri star with a mass of $0.5~\mathrm{M}_\odot$, a radius of $2~\mathrm{R}_\odot$, and an effective temperature of $4000~\mathrm{K}$. In the models, the disk is externally irradiated by UV photons from the interstellar radiation field and a nearby massive O-type star at a distance of $0.1$ pc. We note that the radiation field in the model goes up to $4\times 10^4~\mathrm{G}_0$, which is higher than the radiation field illuminating the two disks presented here ($<10^3~\mathrm{G}_0$).  
The main results of these models are the following:

{\bf{Brighter molecular line emission.}}  
Some molecular lines may be brighter for the irradiated disks because of the higher disk temperatures. In particular, this should occur for CO (and their isotopologues), CN, and HCN. Our detections of $^{12}$CO, $^{13}$CO, and C$^{18}$O are not consistent with this prediction, but this could be due to the cloud contamination discussed in Section \ref{sec:molecular-lines}. In the case of HCN, the brighter line emission in irradiated disks compared to isolated disks seems to be related to the stellar mass and not to the external radiation field. In addition, the measured HCN excitation temperature of $\sim 20$ K is similar to what has been found in isolated disks, suggesting that the HCN emission arises from relatively cold gas with the caveat that the lines could be optically thick. A possible explanation is that we are observing the inner regions of the disk that are currently shielded and no longer affected by the external radiation. But more importantly, it is possible that the external radiation field is just not high enough to produce a significant difference in the line emission. 

{\bf{COMs enhancement.}} 
Chemical models predict that complex organic molecules that are typically frozen out onto dust grains could be observed in the gas-phase in warmer and externally irradiated disks because of their higher temperatures that sublimate these molecules into the gas-phase, in particular in the outer disk. For example, formaldehyde, as a precursor and tracer of COMs, is expected to be enhanced in the outer disk, where the temperature should be higher. However, the observed H$_2$CO radial profiles in the two irradiated disks are centrally peaked, with no additional emission component in the outer disks that would be indicative of ice desorption (see Figure \ref{fig:radial_profiles}). However, we note that the SNR of the detected H$_2$CO lines is low, so it is possible that the current observations are not sensitive enough to detect such a component in the outer disk.

{\bf{Radiation field tracers.}} 
The HCN/HCO$^+$ and CN/HCN line ratios are expected to be larger in irradiated disks compared to isolated disks. HCN/HCO$^+$ ratios larger than one can be indicative of enhanced external radiation, because HCO$^+$ traces the cold, dense regions of the disk, which are smaller for irradiated disks \citep{walsh_molecular_2013}. \citet{factor_alma_2017} measured a HCN/HCO$^+$ ratio of $0.58\pm0.04$ for the 216--0939 disk, which is consistent with the isolated models from \citet{walsh_molecular_2013}. $\mathrm{CN/HCN}>1$ can also be indicative of enhanced radiation fields, similar to what is observed in photodissociation regions (PDRs), where the CN/HCN ratio is found to correlate with the FUV field \citep{fuente_chemical_1993}. This is because FUV photons enhance the abundance of CN in the outer disk, which is more exposed to external radiation \citep{guzman_cyanide_2015,bergner_molecules_2021}. Unfortunately, CN has not yet been observed in these irradiated disks. Future observations of CN lines are needed to verify this prediction. 

\section{Summary} \label{sec:conclusion}
We have presented ALMA observations of the continuum and line emission of several molecules toward two externally FUV-irradiated protoplanetary disk systems around pre-main sequence stars in the outskirts of the Orion Nebula Cluster. In particular, we presented observations of the CO isotopologues, the small organic molecules HCN and H$_2$CO, the carbon chains C$_2$H and $c-$C$_3$H$_2$, and the deuterated species DCN and N$_2$D$^+$, all of which have been previously detected in isolated disks. The main conclusions are the following:

\begin{itemize}
\item The high-angular resolution observations of the dust continuum emission in \mbox{216--0939} shows the presence of a large gap in the inner disk, that is well resolved for the first time in our data. Additionally, the dust emission is asymmetric or eccentric, with the southern side of the disk being $23\%$ brighter than the northern side. We estimate the outer edge of this gap to be around $120-135~\mathrm{au}$, and a disk size of $311.5\pm 14.5~\mathrm{au}$. The high-angular resolution observations of the dust continuum emission allow us to separate the two members in the binary system 253--1536A/B, measure their disk sizes ($239.1\pm14.5~\mathrm{au}$ for A, and $108.7\pm14.5~\mathrm{au}$ for B) and the separation between their edges ($\sim124~\mathrm{au}$). In addition, we do not observe substantial substructure with the current observations, in neither 216--0939 nor 253--1536A/B.
\item We detected the $^{12}$CO $(2-1)$, $^{13}$CO $(2-1)$ and C$^{18}$O $(2-1)$ lines, as well as the HCN $(3-2)$, H$_2$CO $(3-2)$, and C$_2$H~$(3-2)$ lines toward the 216--0939 and 253--1536A/B disks. The CO and CO isotopologue emission is affected by cloud contamination. We estimated the disk-integrated column densities of HCN, H$_2$CO and C$_2$H, assuming optically thin emission, and a range of excitation temperatures, and found values in the range of $10^{13}-10^{14}$~cm$^{-2}$, similar to what is observed in isolated disks.
\item Molecular lines such as $c-$C$_3$H$_2$ $(6_{06}-5_{15})$, $c-$C$_3$H$_2$ $(6_{25}-5_{14})$, $c-$C$_3$H$_2$ $(7_{07}-6_{16})$, \mbox{DCN $(3-2)$} and N$_2$D$^+$ $(3-2)$ were not detected in neither the 216--0939 nor 253--1536A/B disks. 
The cold tracers N$_2$D$^+$ and DCN are expected to be less abundant in the warmer irradiated disks compared to isolated disks, since these molecules are formed more efficiently at low temperatures. However, the estimated upper limits for the disk-integrated fluxes are consistent with detections of these lines in isolated disks. 
\item In general, we do not observe significant differences between the chemistry of isolated and the two irradiated disks presented in this work, based on the observed disk-integrated fluxes and flux ratios for the molecular lines presented here. The differences between the 216--0939 or 253--1536A/B disks and typical T Tauri and Herbig Ae/Be disks found in low-mass star forming regions, seem to be more closely related to the different stellar masses than to the presence of an enhanced external radiation field. 
\item The observed disk integrated fluxes and line ratios in the two irradiated disks presented here are not consistent with chemical model predictions of externally irradiated disks presented by  \cite{walsh_molecular_2013}. However, the 216--0939 and 253--1536A/B disk are irradiated by a weaker FUV field ($<10^3 \mathrm{G}_0$) than the one included in the models ($\sim 10^4~\mathrm{G}_0$). Our results, therefore, suggest that these disks are far enough away from the ONC so that their chemistry is no longer substantially affected, but disks located closer to the stellar cluster may experience stronger chemical effects. 
\end{itemize}

The results presented in this work show that the chemical composition in these moderately irradiated systems is similar to that in isolated disks, which suggests that the assembly of planetary systems and their atmospheres will proceed in a similar manner to that expected in the better studied isolated systems.

Future observations of disks exposed to higher radiation fields are needed to better determine the differences between isolated and externally irradiated disks. In particular, observations of disks closer to the ONC are needed to investigate how the chemistry changes with distance from the ionizing source. In addition,  observations with better sensitivity are needed to detect lines from cold molecular tracers, such as DCN, and determine whether they are indeed less abundant in the warmer irradiated disks compared to isolated disks. 
Finally, chemical models including more  moderate radiation fields are needed to further investigate how the chemistry is affected by an external radiation field.

\section*{Acknowledgments}

We thank the referee for a careful read of the manuscript, and helpful comments that improved the presentation and discussion of the paper.

J.~K.~D.-B. acknowledges support in the form of a studentship from the Science and Technology Facilities Council (grant number ST/X508536/1).

V.V.G gratefully acknowledges support from FONDECYT Regular 1221352, and ANID CATA-BASAL project FB210003. 

C.W.~acknowledges financial support from the Science and Technology Facilities Council and UK Research and Innovation (grant numbers ST/X001016/1 and MR/T040726/1).

L.I.C. acknowledges support from NASA ATP 80NSSC20K0529. L.I.C. also acknowledges support from NSF grant no. AST-2205698, the David and Lucille Packard Foundation, and the Research Corporation for Scientific Advancement Cottrell Scholar Award.

E.A.dlV. acknowledges financial support provided by FONDECYT grant 3200797.

\bibliography{aastex631.bib}{}
\bibliographystyle{aasjournal}

\appendix
\counterwithin{figure}{section}

\setcounter{table}{0}
\renewcommand{\thetable}{A\arabic{table}}

\section{Spectral windows}\label{sec:appendixA}
The spectral settings for Band 6 observations are shown in Table \ref{tab:spectral-setting-app}.

\begin{deluxetable}{@{\extracolsep{10pt}}llcc}
\tablecaption{Spectral settings.}
\tablewidth{0pt}
\tablehead{
\colhead{Molecule} & \multicolumn{1}{c}{Line}   &  \colhead{Resolution [km s$^{-1}$]} & \colhead{Bandwidth [MHz]}\\
}
\startdata
\hline
\hline
\multicolumn{4}{c}{220 GHz Setting}\\
\hline
\hline
DCN & $3-2$ & $0.195$ & $58.59$ \\
$c-$C$_3$H$_2$ & $6_{06}-5_{15}$ & $0.194$ & $58.59$ \\
H$_2$CO & $3_{22}-2_{21}$ 	&   $0.194$  & $58.59$ \\
C$^{18}$O & $2-1$	            &   $0.193$  & $58.59$ \\
$^{13}$CO  & $2-1$	            &   $0.192$  & $58.59$ \\
$^{12}$CO & $2-1$ 		        &   $0.184$  & $117.19$ \\
N$_2$D$^+$ & $3-2$    	        &   $0.183$  & $117.19$ \\
\hline
\hline
\multicolumn{4}{c}{265 GHz Setting}\\
\hline
C$_2$H & $3-2$    	            &   $0.161$ & $234.38$ \\
$c-$C$_3$H$_2$ & $6_{25}-5_{14}$ &   $0.168$ & $58.59$ \\
           & $7_{07}-6_{16}$ &   $0.168$ & $58.59$ \\
HCN & $3-2$		                &   $0.159$ & $234.38$ \\
\hline
\enddata
\tablecomments{Spectral settings of the molecular line transitions. It includes the bandwidth for each spectral window.
}
\label{tab:spectral-setting-app}
\end{deluxetable}

\section{Channel Maps}\label{sec:appendixB}
\begin{figure*}[ht]
\centering
\includegraphics[width=18cm]{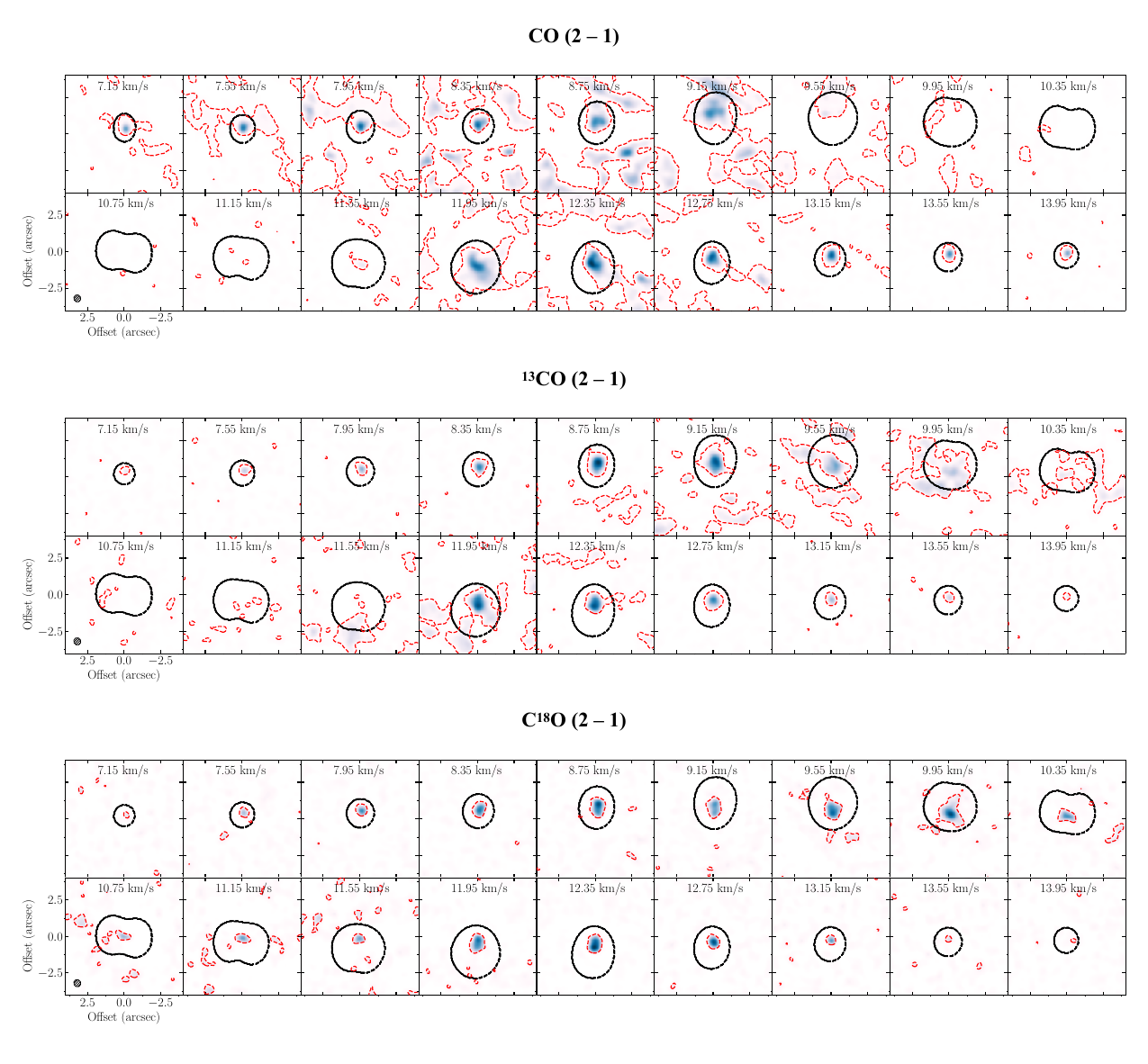}
\caption{
Channel maps for the CO~($2-1$), $^{13}$CO~($2-1$), and C$^{18}$O~($2-1$) lines in 216--0939 (systemic velocity of $10.75$ km/s). The Keplerian mask is shown in black. 
Red contours correspond to the $3\sigma$ level of the emission. 
The synthesized beam is shown in the bottom left panel.
\label{fig:appA}} 
\end{figure*}

\begin{figure*}[ht]
\centering
\includegraphics[width=18cm]{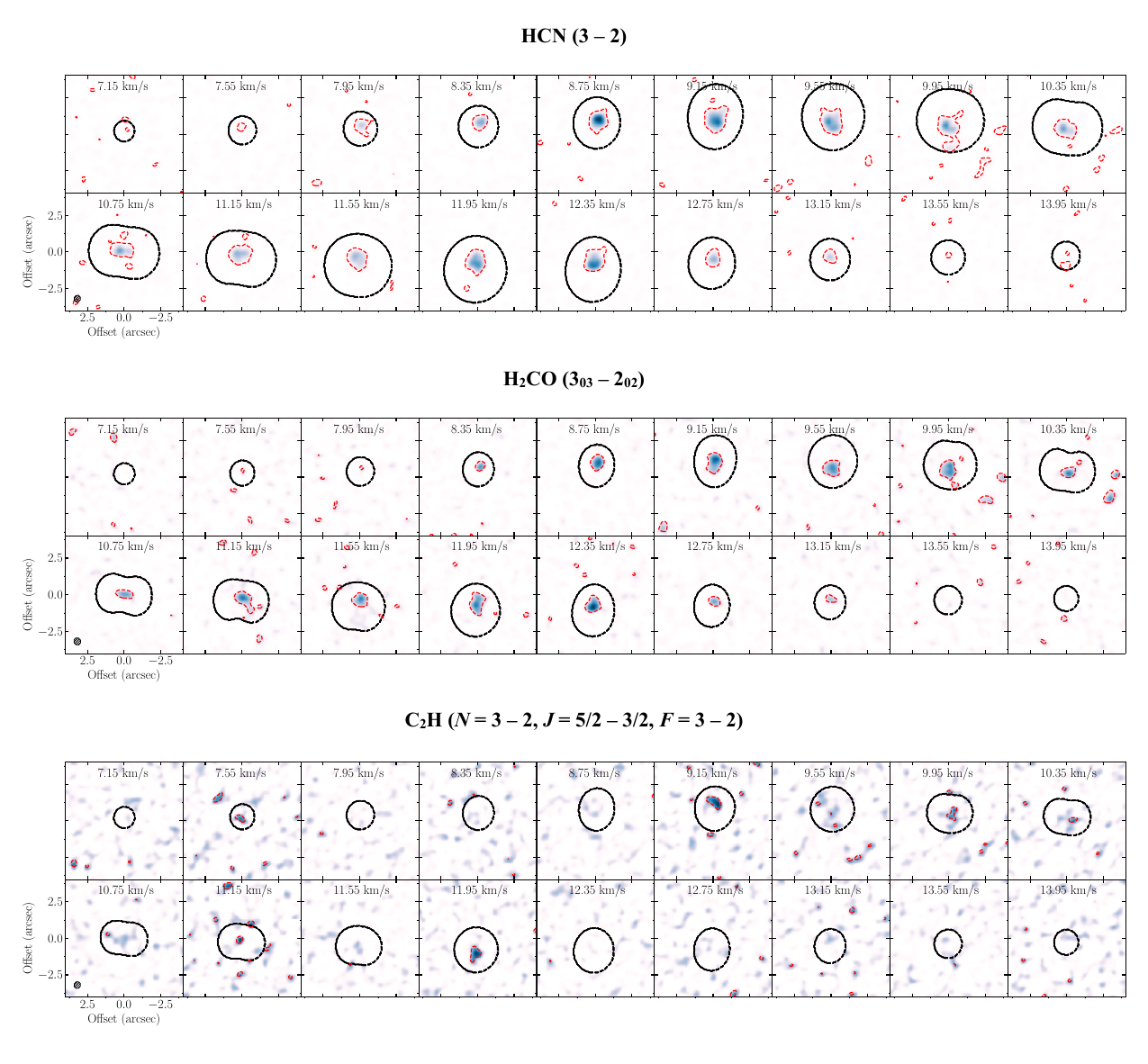}
\caption{
Same as \ref{fig:appA} but for HCN~($3-2$), H$_2$CO~($3-2$), and C$_2$H~($3-2$) lines in 216--0939.}
\label{fig:appA-2}
\end{figure*}

\begin{figure*}[ht]
\centering
\includegraphics[width=18cm]{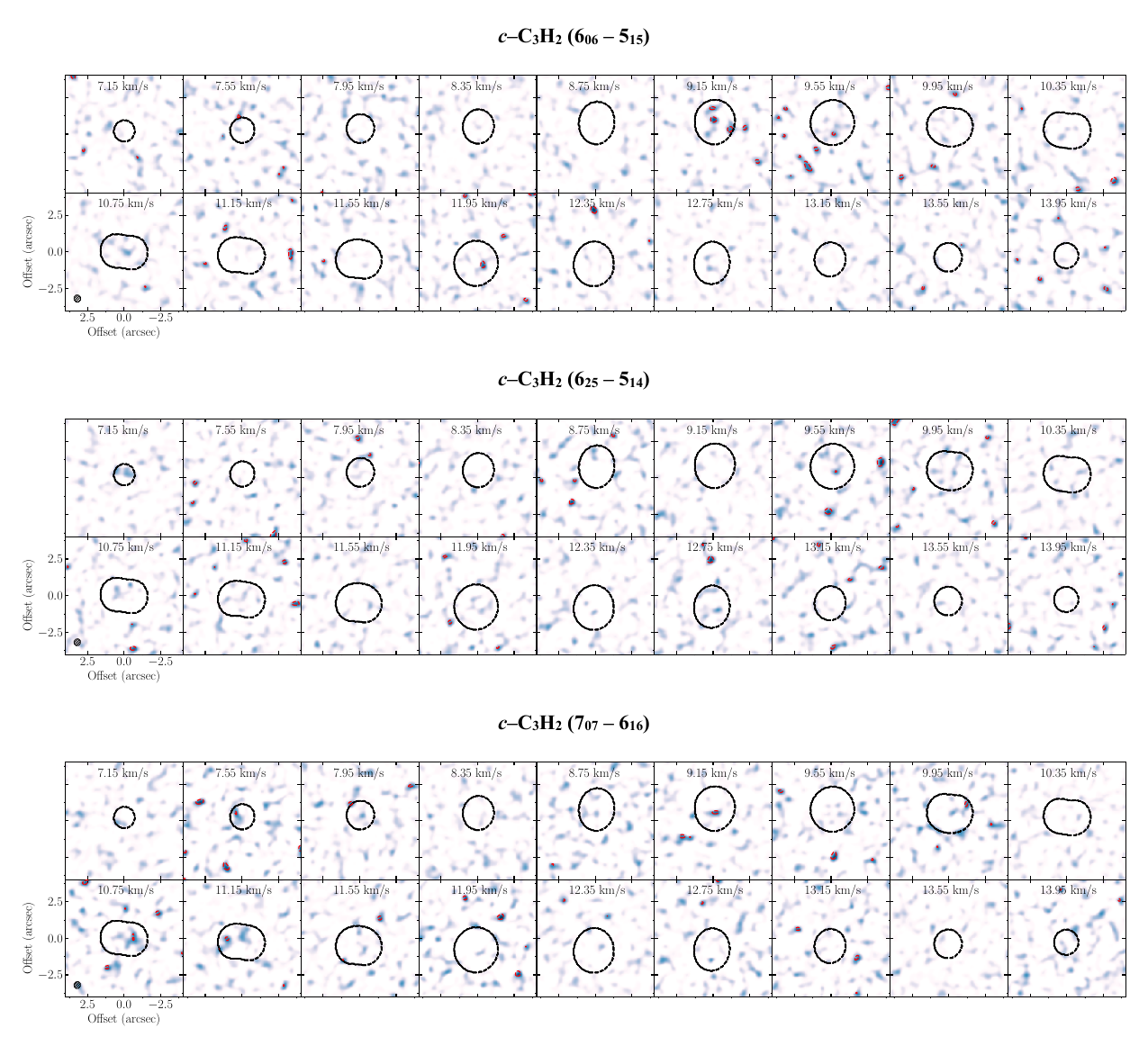}
\caption{
Same as \ref{fig:appA} but for $c-$C$_3$H$_2$~($6_{06}-5_{15}$), $c-$C$_3$H$_2$~($6_{25}-5_{14}$), and $c-$C$_3$H$_2$~($7_{07}-6_{16}$) lines in 216--0939.}
\label{fig:appA-3}
\end{figure*}

\begin{figure*}[ht]
\centering
\includegraphics[width=18cm]{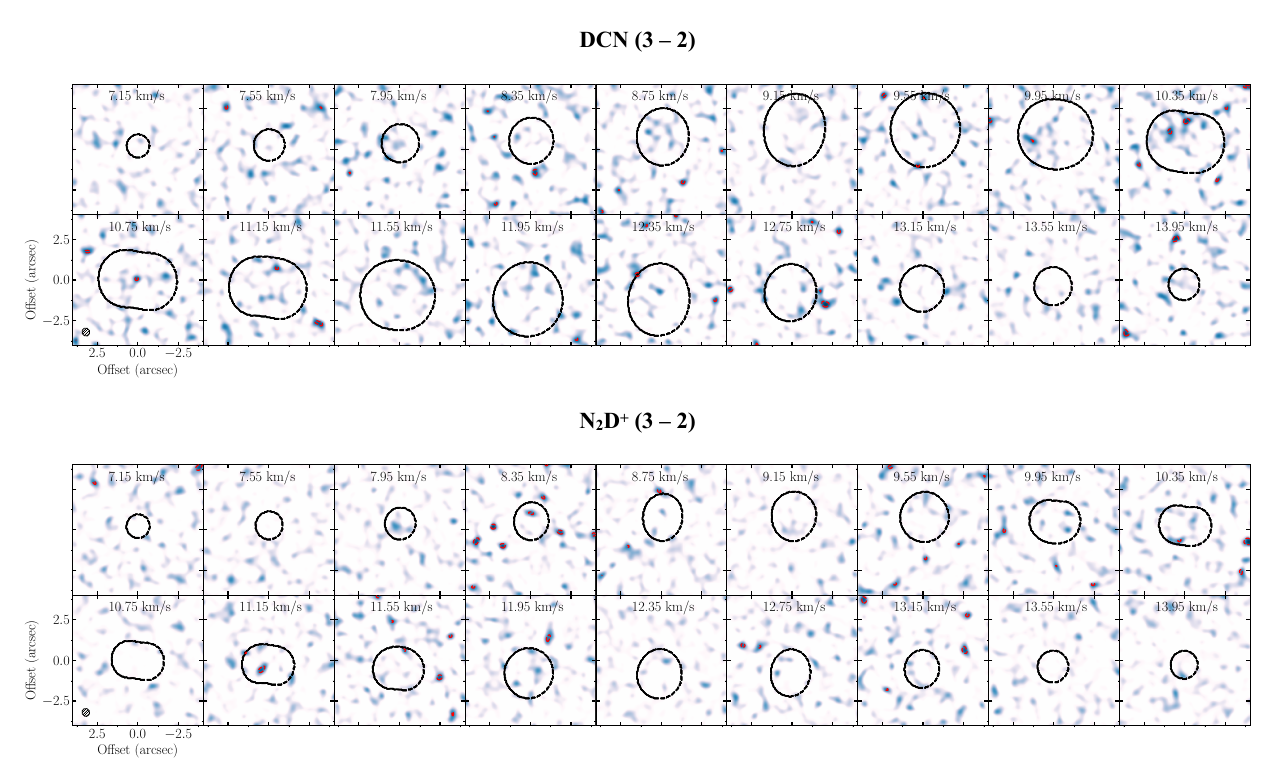}
\caption{
Same as \ref{fig:appA} but for DCN~($3-2$), and N$_2$D$^+$~($3-2$) lines in 216--0939.}
\label{fig:appA-4}
\end{figure*}

\begin{figure*}[ht]
\centering
\includegraphics[width=18cm]{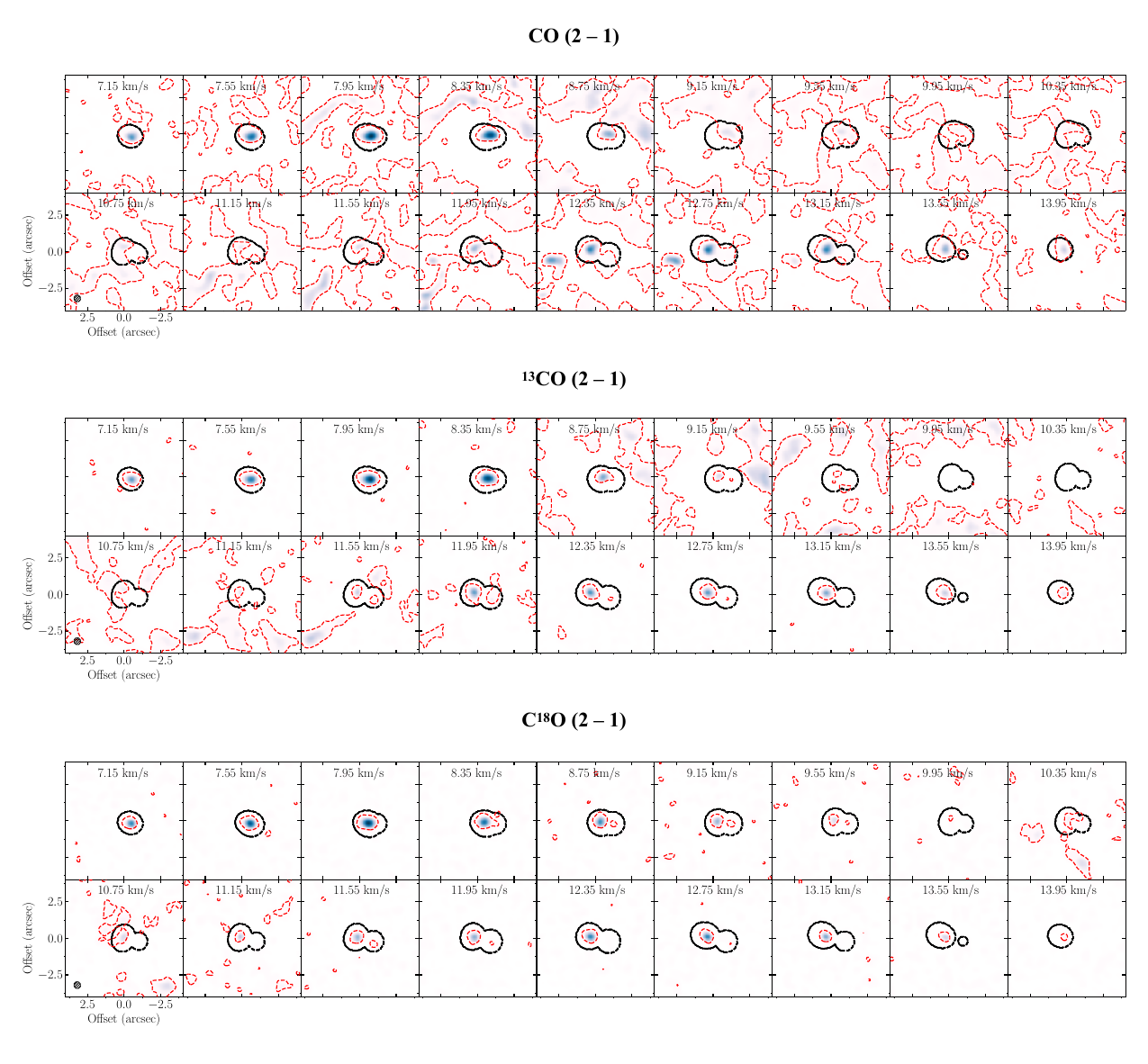}
\caption{
Same as \ref{fig:appA} but for CO~($2-1$), $^{13}$CO~($2-1$), and C$^{18}$O~($2-1$) lines in 253--1536 (systemic velocity of \mbox{$10.55$ km/s}).}
\label{fig:appA-5}
\end{figure*}

\begin{figure*}[ht]
\centering
\includegraphics[width=18cm]{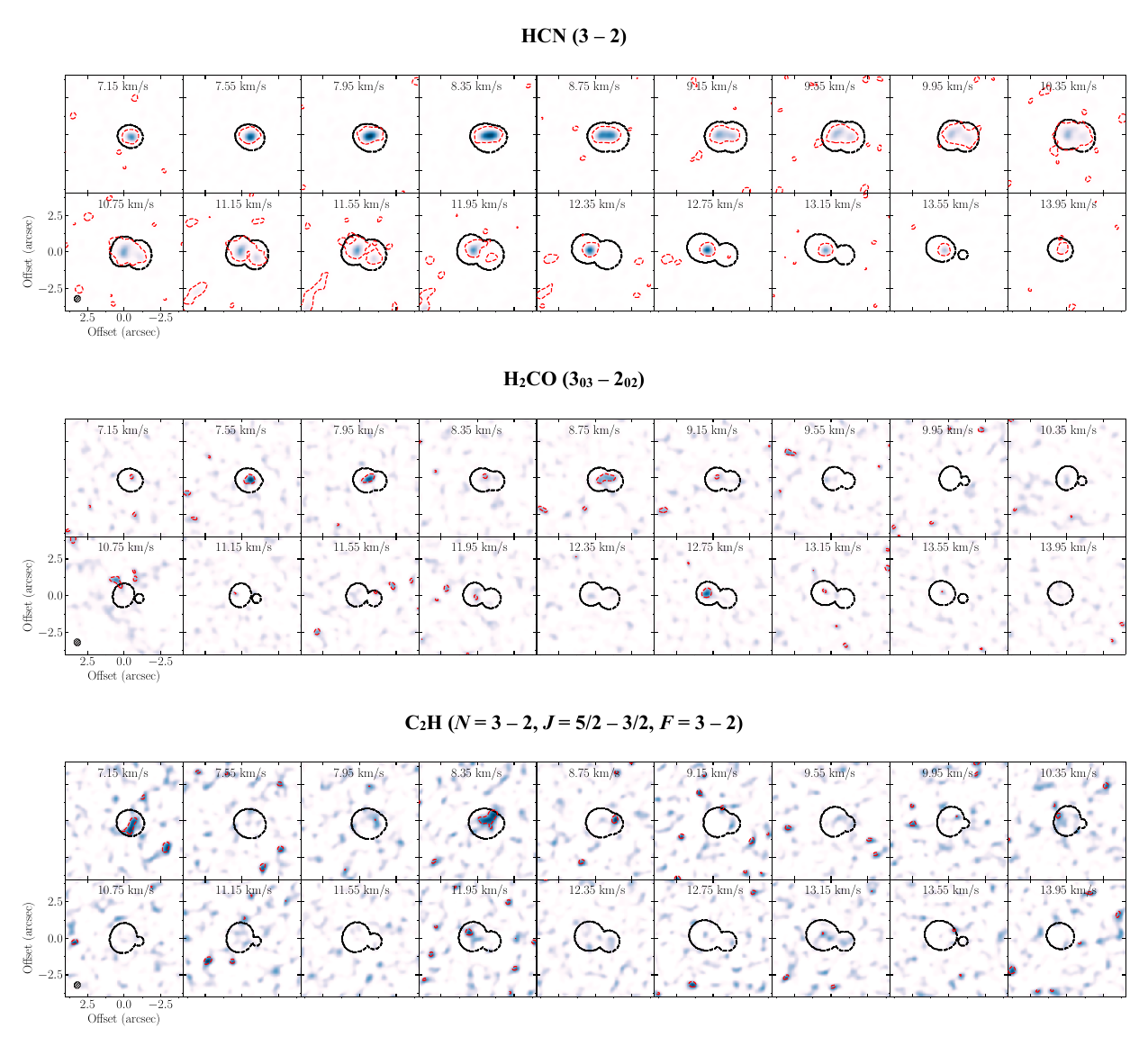}
\caption{
Same as \ref{fig:appA} but for HCN~($3-2$), H$_2$CO~($3-2$), and C$_2$H~($3-2$) lines in 253--1536.}
\label{fig:appA-6}
\end{figure*}

\begin{figure*}[ht]
\centering
\includegraphics[width=18cm]{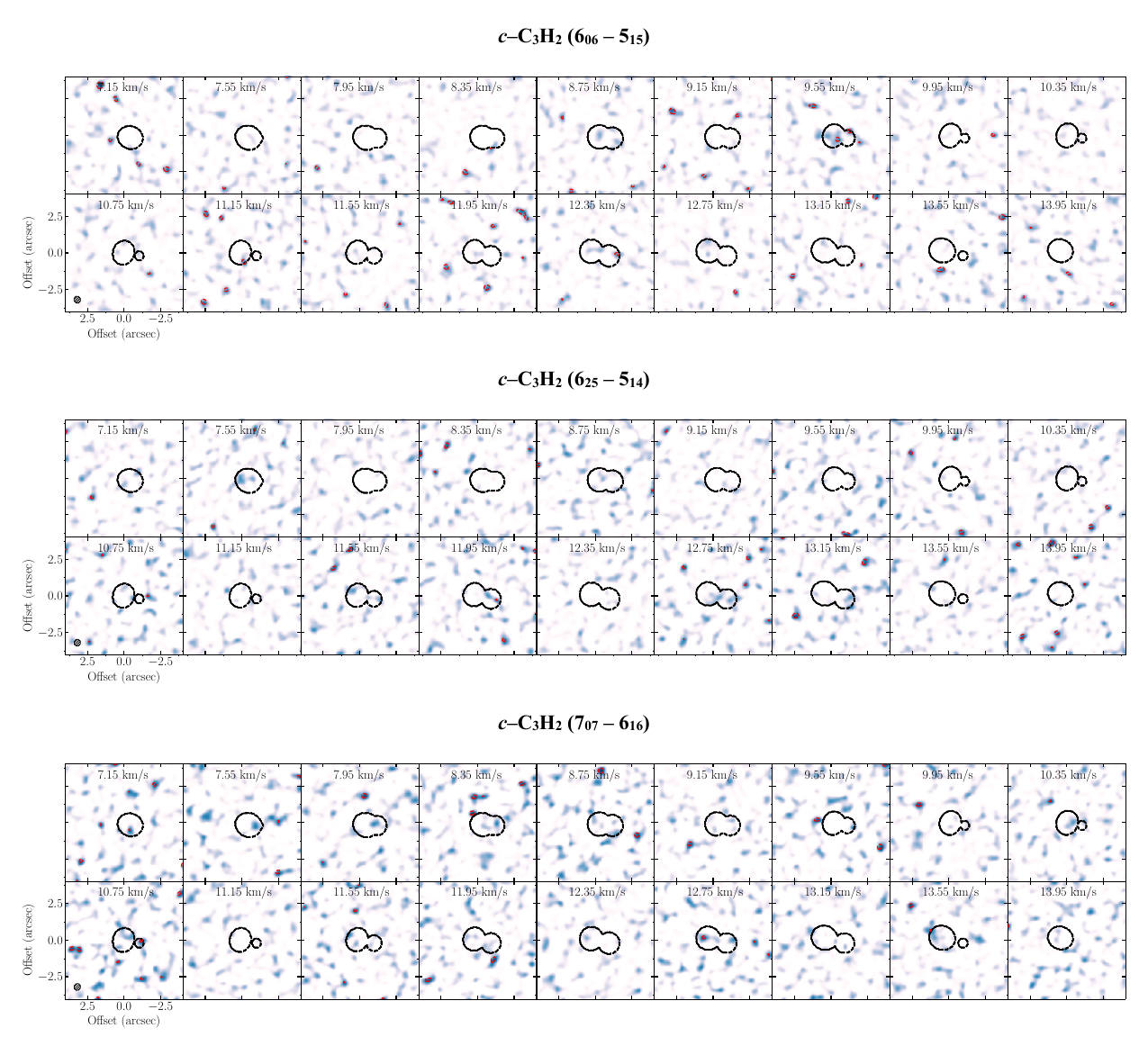}
\caption{
Same as \ref{fig:appA} but for $c-$C$_3$H$_2$~($6_{06}-5_{15}$), $c-$C$_3$H$_2$~($6_{25}-5_{14}$), and $c-$C$_3$H$_2$~($7_{07}-6_{16}$) lines in 253--1536.}
\label{fig:appA-7}
\end{figure*}

\begin{figure*}[ht]
\centering
\includegraphics[width=18cm]{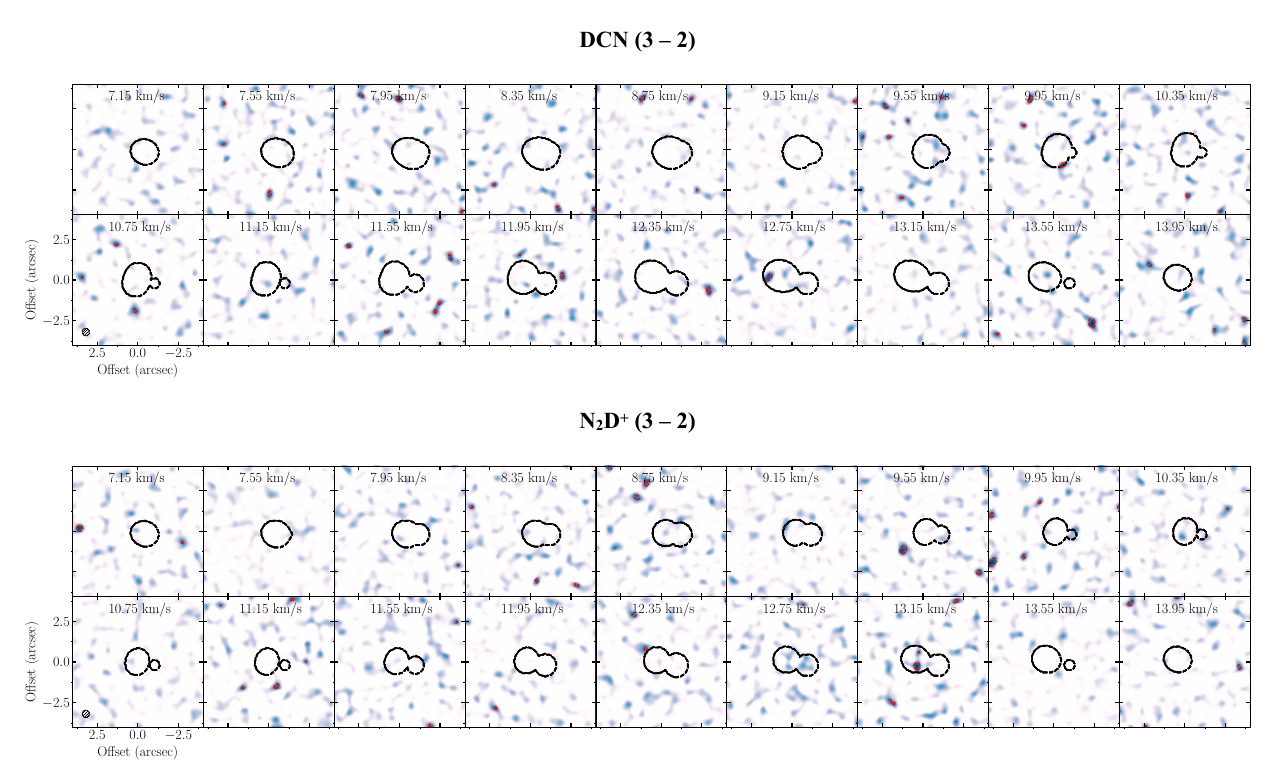}
\caption{
Same as \ref{fig:appA} but for DCN~($3-2$), and N$_2$D$^+$~($3-2$) lines in 253--1536.}
\label{fig:appA-8}
\end{figure*}

The channel maps of the observed molecular lines are shown in Figures \ref{fig:appA}, \ref{fig:appA-2}, \ref{fig:appA-3}, and \ref{fig:appA-4} for the 216--0939 disk, and in Figures \ref{fig:appA-5}, \ref{fig:appA-6}, \ref{fig:appA-7}, and \ref{fig:appA-8} for the 253--1536A/B system.

\section{Matched filter results}\label{sec:appendixC}
\begin{figure*}[ht]
\centering
\includegraphics[width=16cm]{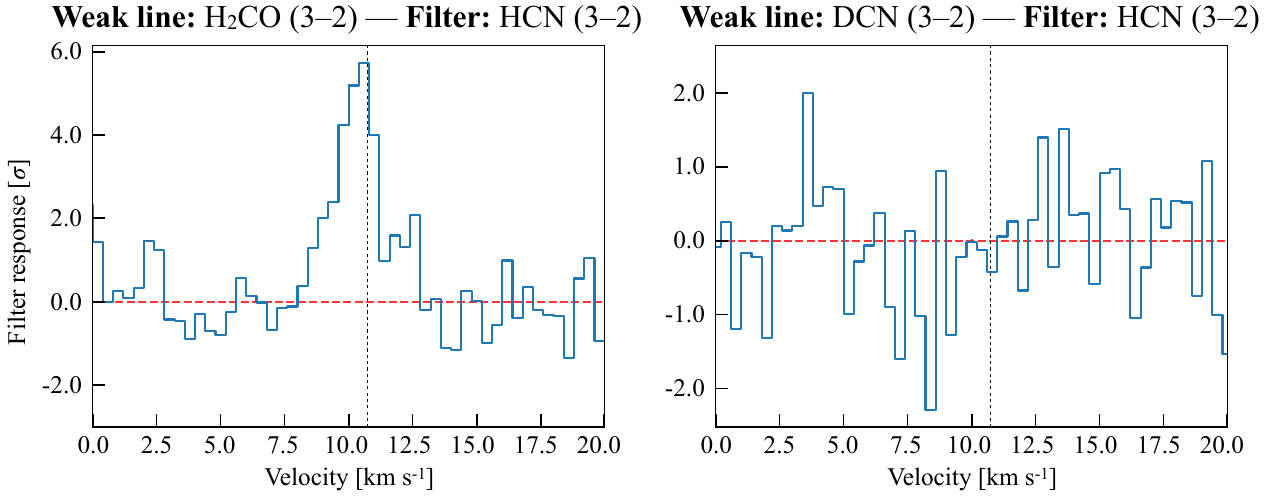}
\caption{Filter response spectra for H$_2$CO $(3-2)$ (\textsl{left}) and DCN $(3-2)$ (\textsl{right}) in the 216--0939 disk. Both impulse responses are obtained using HCN~$(3-2)$ as the filter.
\label{fig:visible_H2CO_DCN}} 
\end{figure*}

\begin{figure*}[ht]
\centering
\includegraphics[width=16cm]{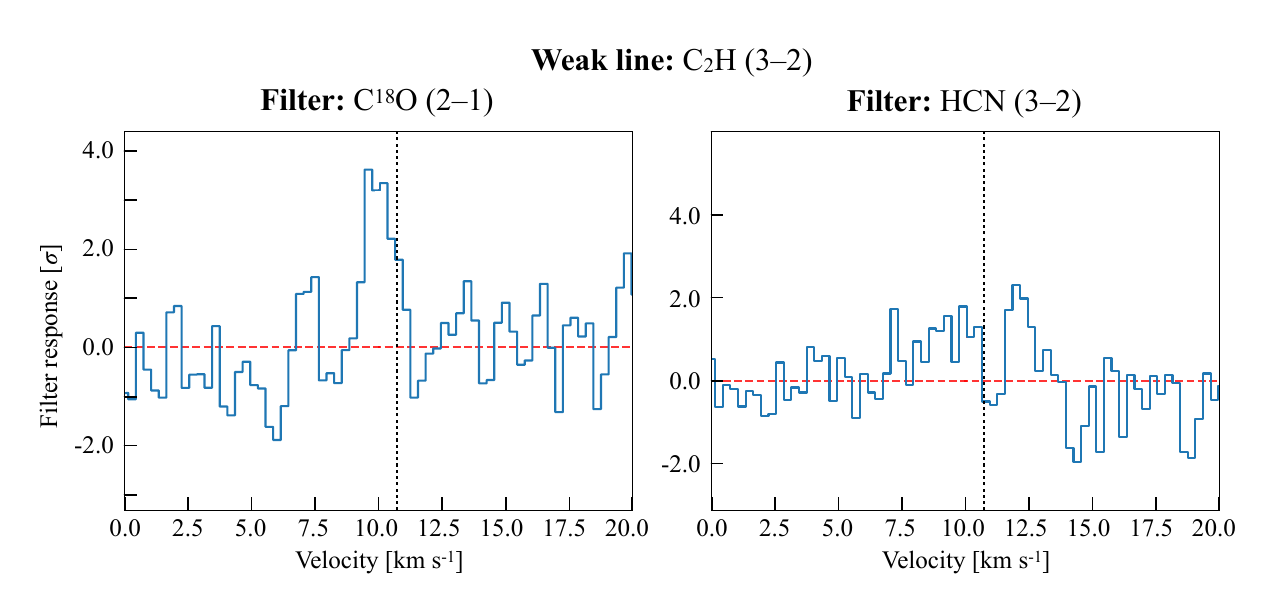}
\caption{
Filter response spectra for C$_2$H~$(3-2)$ toward the 216--0939 disk, obtained using C$^{18}$O $(2-1)$ \textsl{(left)} and HCN~$(3-2)$ \textsl{(right)} as filters.
\label{fig:visible_C2H}} 
\end{figure*}

Figs.~\ref{fig:visible_H2CO_DCN} and \ref{fig:visible_C2H} show examples of the filter response spectra from the matched–filter technique \citep[\texttt{VISIBLE};][]{loomis_detecting_2018}. The first figure shows the difference in the response for a bright detected line (H$_2$CO) compared to a weak non-detected line (DCN), using the HCN line as a filter in both cases. The second figure shows the difference in the response for C$_2$H when using different filters. A clear peak is seen in the response when the C$^{18}$O line is used as a filter, suggesting the C$_2$H line is detected, and a less clear detection is seen when instead HCN is used as a filter. This suggests that the distribution of C$_2$H is likely more similar to that of C$^{18}$O than that of HCN.

\section{Disk integrated spectra}\label{sec:app_spectra}

\begin{figure}[b!]
\centering
\includegraphics[width=0.7\textwidth]{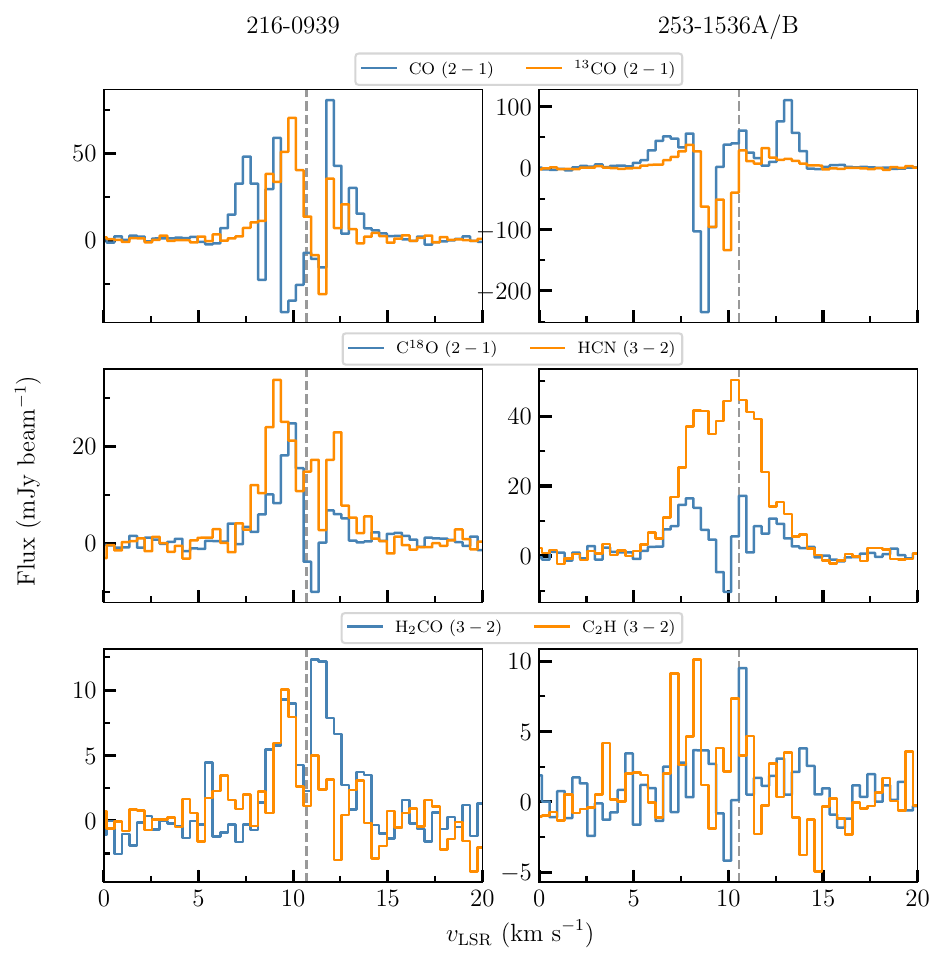}
\caption{
Disk integrated spectra for 216--0939 \textsl{(left panels)} and for 253--1536A/B \textsl{(right panels)}.
\textsl{Upper panels:} CO and $^{13}$CO. \textsl{Central panels:} C$^{18}$O) and HCN. \textsl{Lower panels:} H$_2$CO and C$_2$H.}
\label{fig:spectrum}
\end{figure}

Figure~\ref{fig:spectrum} shows the disk-integrated spectra for the detected lines in 216--0939 and 253--1536A/B. These spectra are obtained by adding all the emission inside the Keplerian mask in each channel. We note that the disk-integrated spectra of the 253--1536A/B system includes emission from both binary members. The CO isotopologues and HCN emission show a double peaked profile, typical of Keplerian rotation of an inclined disk, in both sources. However, the CO isotopologue lines are heavily affected by absorption from the molecular cloud, in particular in the central channels. For molecular lines such as H$_2$CO and C$_2$H, it is more difficult to distinguish the Keplerian rotation associated with the disk, because these lines are faint.

\section{Deviations from Keplerian rotation in 253--1536A/B}\label{sec:app_253}

\begin{figure}[h!]
\centering
\includegraphics[width=8.5cm]{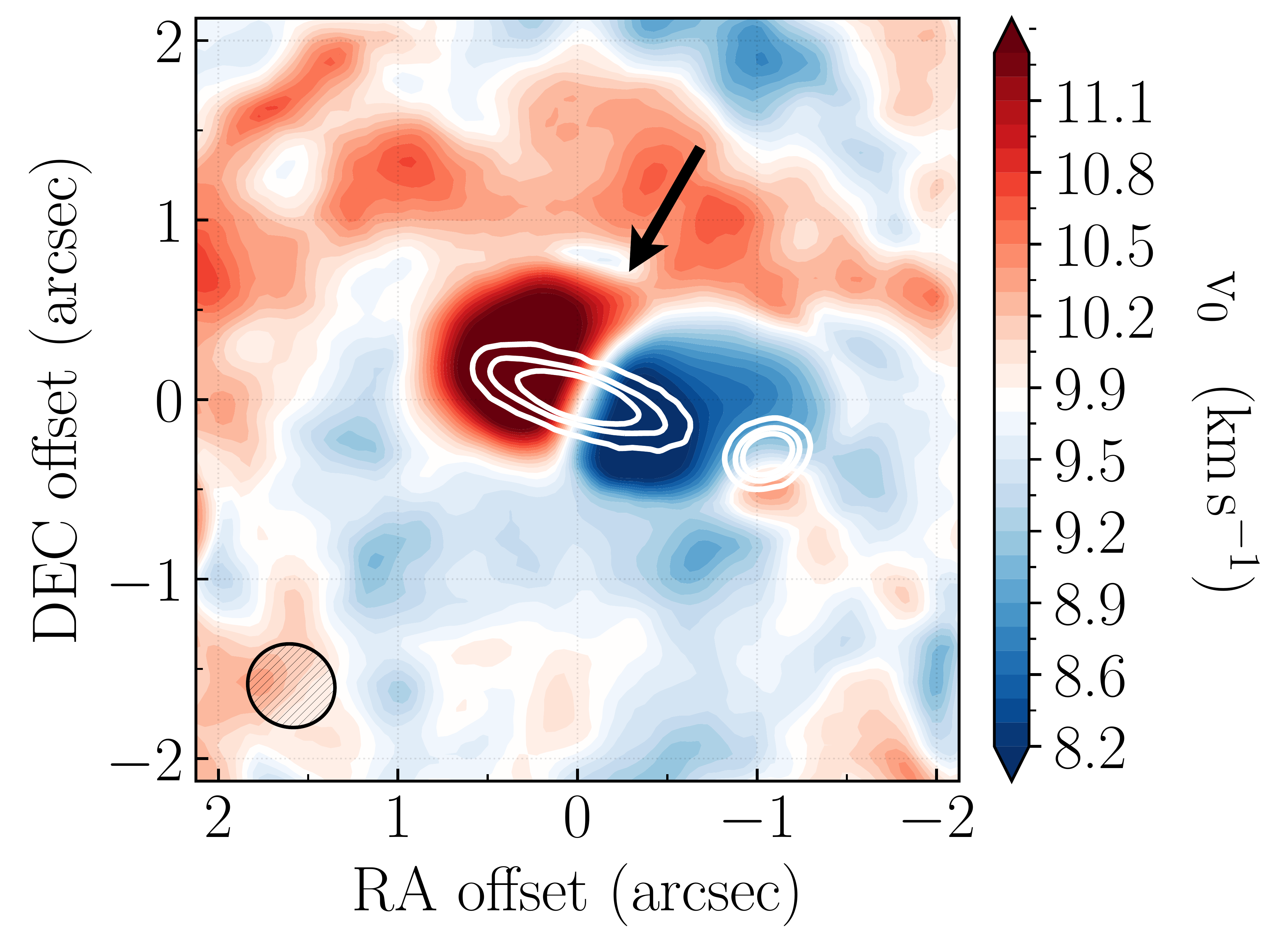}
\caption{
First--moment of the CO $(2-1)$ line emission for 253--1536A/B. 
Colors represent red- and blue-shifted parts of the cube relative to the source velocity. The beam size is shown in the bottom left corner of the figure. 
White contours correspond to $2$, $15$, and $30\sigma$ of the dust continuum emission of the sources. The black arrow represents a tentative deviation from Keplerian rotation.
\label{fig:CO_mom1}} 
\end{figure}

Figure~\ref{fig:CO_mom1} shows the first-moment map of the CO $(2-1)$ line emission for the 253--1536A/B system. The moment was obtained using the \texttt{bettermoments} Python package \citep{teague_evidence_2018, teague_eddy_2019}, which collapses the emission cube and estimates the intensity weighted average velocity in each pixel. The figure shows the Keplerian rotation of the disk, mainly for the 253--1536A member. In addition, a tentative deviation from Keplerian rotation is seen towards the northern side of the 253--1536A star, which is indicated by the black arrow.  

\end{document}